\UseRawInputEncoding
\documentclass[showpacs,twocolumn,superscriptaddress,nofootinbib]{revtex4-1} 

\usepackage{hyperref}
\hypersetup{colorlinks=true,linkcolor=blue,citecolor=cyan}
\usepackage{graphicx}
\usepackage{xcolor, soul}
\sethlcolor{green}
\usepackage{dcolumn}
\usepackage{bm}
\usepackage{color}
\usepackage{enumitem}
\usepackage{mathrsfs}
\usepackage{amsmath}
\usepackage{amssymb}
\usepackage{cleveref}
\usepackage{orcidlink}

\crefname{equation}{Eqn.}{Eqns.}
\crefname{figure}{Fig.}{Figs.}
\crefname{section}{Sec.}{Sec.}
\crefname{table}{Table}{Tables}
\usepackage{physics}
\usepackage{multirow}

\setstcolor{red}

\begin{document}


\title{Constraining ModMax Black Holes with EHT and GRAVITY Observations: Optical Signatures and Accretion Disk Properties}

\author{Tursunali Xamidov 
}
\email{xamidovtursunali@gmail.com}
\affiliation{Institute of Fundamental and Applied Research, National Research University TIIAME, Kori Niyoziy 39, Tashkent 100000, Uzbekistan} 
\affiliation{Institute for Theoretical Physics \& Cosmology, Zhejiang University of Technology, Hangzhou 310023, China}

\author{Mirjavoxir Mirkhaydarov 
}
\email{mirxaydarovmirjavohir@gmail.com}
\affiliation{Institute of Fundamental and Applied Research, National Research University TIIAME, Kori Niyoziy 39, Tashkent 100000, Uzbekistan}

\author{Sanjar Shaymatov 
}
\email{sanjar@astrin.uz}
\affiliation{Institute of Fundamental and Applied Research, National Research University TIIAME, Kori Niyoziy 39, Tashkent 100000, Uzbekistan}
\affiliation{University of Tashkent for Applied Sciences, Str. Gavhar 1, Tashkent 100149, Uzbekistan}
 \author{Pankaj Sheoran 
 }
\email{pankaj.sheoran@vit.ac.in}
\affiliation{Department of Physics, School of Advanced Sciences, Vellore Institute of Technology, Tiruvalam Rd, Katpadi, Vellore, Tamil Nadu 632014, India}

\author{Chengxun Yuan}
   \email{yuancx@hit.edu.cn}
\affiliation{School of Physics, Harbin Institute of Technology, Harbin 150001, People’s Republic of China}

\date{\today}
\begin{abstract}

In this work, we study photon propagation and the radiative properties of a thin accretion disk around a ModMax black hole and examine the effects of the charge $Q$ (i.e. $Q^2 = Q^2_e +Q^2_m$ is the total
dyonic charge, with $Q_e$ and $Q_m$ denoting the electric and
magnetic charges, respectively) and the nonlinearity parameter $v$. We determine the event-horizon, photon-sphere, and shadow radii and find that increasing $Q$ decreases these radii, whereas increasing $v$ shifts them toward their Schwarzschild values. Using the EHT shadow measurements of M87$^\star$ and Sgr~A$^\star$, together with the available mass and distance measurements, we perform an MCMC analysis to constrain the ModMax parameters. The strongest upper limits are found to be $Q<0.391$ and $v<4.153$ at the 95\% credible level. We also investigate a Novikov--Thorne thin accretion disk and generate simulated disk images using backward ray tracing. The observed flux increases with $Q$, while increasing $v$ produces a slight decrease in the disk brightness. These results show how the ModMax parameters affect the black-hole shadow and thin-disk emission and provide observational constraints on $Q$ and $v$.
\end{abstract}

\maketitle

\section{Introduction}
\label{introduction}
The quest to understand gravity in its most extreme form has taken black hole (BH) physics to new heights of modern astrophysical research. While general relativity (GR) has been tested with high precision in the weak-field regime, the strong gravitational fields in the close vicinity of BHs offer a unique testing ground where deviations from GR, if they exist, would become most significant \cite{Ni:2016uik,Sheoran:2017dwb,Gralla:2020srx,Gralla:2019xty,Flores-Alfonso:2020euz, Bandos:2020jsw, Siahaan:2024ioa, Pantig:2022gih}. 
This has motivated many research works \cite{Bandos:2020jsw, Barrientos:2022bzm, EslamPanah:2024gxx, Sekhmani:2025epe} to explore extensions and alternatives to GR, as well as novel matter field configurations. These works aim to identify robust observational signatures that can differentiate between various competing theoretical models. Among the various probes to study the strong gravity regimes, BH shadows \cite{Tsukamoto:2014tja,Wei:2019pjf,Gralla:2019xty,Psaltis:2020ctj} and the properties of accretion disks \cite{Jiang:2014loa,Jiang:2015dla,Ni:2016uik,Wang:2018bbr,Tripathi:2018bbu,Tripathi:2021rqs} have emerged as particularly powerful tools because they are directly accessible to current experiments such as the Event Horizon Telescope (EHT) \cite{EventHorizonTelescope:2019dse,EventHorizonTelescope:2019ggy,EventHorizonTelescope:2022wkp,EventHorizonTelescope:2022xqj} and the GRAVITY collaboration \cite{2018Natur.563..657G,GRAVITY:2020gka}. However, the present observations impose crucial constraints on strong-field gravity; the next-generation missions like the ngEHT and BHEX are expected to probe the strong gravity regime with greater depth and to update the characterization of the near-horizon limits of BHs \cite{Johnson:2023ynn,Uniyal:2025uvc}.

In this context, the ModMax theory of nonlinear electrodynamics has gained significant attention. This theory is unique as it is the one-parameter generalization of Maxwell's theory that preserves both conformal invariance and electromagnetic duality while continuously reducing to Maxwell electrodynamics in the appropriate limit \cite{Bandos:2020jsw, Kosyakov:2020wxv, Flores-Alfonso:2020euz, Sorokin:2021tge}. This new parameter acts as a charge screening factor that modifies the effective charge of BHs. Hence, it alters their geometric and thermodynamic properties and observational signatures \cite{Barrientos:2022bzm, EslamPanah:2024gxx, Siahaan:2024ioa}. The ModMax theory is interesting not because it’s the simple generalization of Maxwell’s theory or due to its theoretical consistency but because of its potential to resolve issues present in other nonlinear electrodynamics, such as superluminal propagation, making it a physically viable candidate for beyond-Maxwell physics \cite{Kosyakov:2020wxv,Sorokin:2021tge, Russo:2024xnh}.

In the last few years, there has been considerable interest in a distinctive class of nonlinear electrodynamics known as modified Maxwell, or ModMax, theory \cite{Sorokin:2021tge,Siahaan:2024ioa}. Besides the motivation stated in the last paragraph, another motivation for studying nonlinear electrodynamical BH models stems from their ability to avoid the field singularities that are considered one of the shortcomings of classical Maxwell theory-inspired BH models. Hence, ModMax BH models are considered to offer a more complete description of electromagnetic phenomena in extreme gravity regimes \cite{Flores-Alfonso:2020euz, Siahaan:2024ioa}. Following these motivations, a comprehensive body of literature has emerged investigating various aspects of ModMax-inspired BH solutions, ranging from thermodynamical properties and particle dynamics to optical signatures \cite{EslamPanah:2024gxx, EslamPanah:2025bfh, Barrientos:2022bzm, Guzman-Herrera:2023zsv, Pantig:2022gih, Amirabi:2022hsv, Shahzad:2024cvs, Al-Badawi:2025coy, Sekhmani:2025epe}. In addition, the charge-screening effect in ModMax theory has been studied in detail \cite{Flores-Alfonso:2020euz, Kosyakov:2020wxv}, wherein the nonlinear parameter introduces an exponential factor that effectively modifies the observed charge of the BH. Another interesting feature of the exact static Einstein-ModMax theory BHs is that they retain a form similar to the classical Reissner-Nordstr\"{o}m solution, with the ModMax parameter simply renormalizing the charge term \cite{Flores-Alfonso:2020euz, Siahaan:2024ioa}.

Additionally, various works have investigated ModMax BHs in a variety of modified gravity frameworks, including dRGT massive gravity \cite{Hassanabadi:2026kti}, \(F(R)\) gravity \cite{EslamPanah:2024gxx, EslamPanah:2025bfh}, and Einstein-Gauss-Bonnet theory \cite{Amirabi:2022hsv}. In \cite{Shahzad:2024cvs, EslamPanah:2025bfh}, the authors analyzed the thermodynamical aspects of the theory, such as phase transitions and Joule-Thomson expansion and showed altered critical phenomena and stability conditions compared to their classical Reissner-Nordstr\"{o}m counterparts. The optical properties of ModMax BHs, such as shadows, gravitational lensing, and quasinormal modes, have also been extensively studied in \cite{Pantig:2022gih, Al-Badawi:2025coy, Jafarzade:2024zqq, EslamPanah:2024gxx}. Particle dynamics around these BHs have also been investigated in the context of geodesic motion, quasi-periodic oscillations, and accretion disks in \cite{Guzman-Herrera:2023zsv, Hoshimov:2025tdx, Karshiboev:2024zyn, Siahaan:2024ioa}. However, it is important to note here that the study of accretion disks is very limited and does not sufficiently address their observational aspects in much detail, which motivates us to investigate the ModMax BHs and bridge this gap. All these works have collectively demonstrated that the nonlinearity parameter of ModMax BHs leaves distinctive imprints across multiple observational channels \cite{EslamPanah:2024gxx, Pantig:2022gih, Al-Badawi:2025coy, Sekhmani:2025epe}.

The BH shadow observations and accretion disk physics together offer a particularly powerful testbed for gravity theories beyond GR. The EHT has provided the first direct images of the supermassive BHs M87$^\star$ and Sgr~A$^\star$, showing the characteristic photon ring and shadow. It is shown in \cite{Al-Badawi:2025coy, Pantig:2022gih, Jafarzade:2024zqq, EslamPanah:2025bfh,Sharipov2026ChJPh} that the size and shape of these shadows are highly sensitive to the underlying BH spacetime. Simultaneously, the GRAVITY collaboration has measured the orbits of stars moving around Sgr~A$^\star$ with very high accuracy and provided the independent constraints on the mass and gravitational potential of the Sgr~A$^\star$ \cite{Siahaan:2024ioa, EslamPanah:2024gxx, Pantig:2022gih}. These observations from EHT and GRAVITY collaborations probe different regimes of the BH spacetime—the strong-field, near-horizon region via the shadow, and the intermediate-field region via stellar dynamics and hence making them ideal for constraining deviations from GR \cite{Bandos:2020jsw, Flores-Alfonso:2020euz}. In this work, we use this complementarity by combining both types of observations while also making use of the independent observations to restrict the ModMax BH parameters separately, so allowing us to impose tighter constraints on the ModMax parameter.

The study of thin accretion disks around BHs provides an important linkage between the theoretical predictions of gravity theories beyond GR and observable electromagnetic signatures. The radiant flux, temperature profile, and emission spectrum of the accretion disks are determined using the underlying BH spacetime metric through the geodesic equations describing the motion of particles in the disk \cite{Guzman-Herrera:2023zsv, Hoshimov:2025tdx, Karshiboev:2024zyn, Siahaan:2024ioa}. Recent observations by the EHT have opened new opportunities to study accretion flows around compact objects. In this context, Sgr A$^{\star}$ provides a valuable astrophysical laboratory for exploring BH accretion and testing gravity in the strong-field regime \cite{Nucita07,you2024,Ghez05,Ghez00,Imdiker23CQG,Vagnozzi23EHT,Afrin23ApJ}. These developments have motivated extensive studies of accretion disks in different gravity models, offering further insight into the properties of BHs and their surrounding spacetime \cite{Shaymatov2023,Boshkayev:2020kle,Collodel_2021,Alloqulov24CPC,Boshkayev21PRD,Hu2023,Alloqulov24EPJP,Cui24,Mirkhaydarov2025CPC,Cai_2025,Nozari2025b,Igata2025,XamidovAccretion2025,Sharipov:2026syz}. For ModMax BHs, the nonlinearity parameter modifies the effective gravitational potential. Hence, influence the location of the innermost stable circular orbit (ISCO) \cite{Dadhich22a,Dadhich22IJMPD}, the disk's inner edge, and hence the entire emission profile \cite{Flores-Alfonso:2020euz, Siahaan:2024ioa, Hoshimov:2025tdx}. By analyzing these observable quantities and comparing them with already obtained constraints, one can assess whether ModMax BHs are consistent with the observed phenomenology of astrophysical BHs, such as the spectral energy distribution of Sgr A* and M87* \cite{Al-Badawi:2025coy, Guzman-Herrera:2023zsv, Pantig:2022gih, Hoshimov:2025tdx}. This approach not only tests the viability of the ModMax theory but also provides a framework for distinguishing it from other modified gravity scenarios.

Despite this much work done by the researchers recently on ModMax BHs, a critical gap remains: a more in-depth, simultaneous analysis that tightly constrains the ModMax parameter using the most recent observations from EHT and GRAVITY, combined observations of BH shadows (EHT) and stellar orbits (GRAVITY), and that rigorously connects these constraints to the physical properties of thin accretion disks. While individual studies have either constrained parameters from shadows or studied accretion disks in isolation, they often do so using simplified models or older data sets \cite{Pantig:2022gih, Guzman-Herrera:2023zsv, Hoshimov:2025tdx}. In addition to the above, the physical interpretation of the accretion disk's properties, such as its flux, temperature, and efficiency, has not been analyzed in depth. In this work we fill this gap by providing a unified framework that bridges the gap between theoretical predictions and observational data.

In addition to our goal of constraining the ModMax parameter, our study contributes to the effort of using BH observations as a probe of fundamental physics. The nonlinear parameter in ModMax electrodynamics can be interpreted as an effective coupling constant that encodes the strength of nonlinear electromagnetic interactions in the strong-field regime \cite{Sorokin:2021tge, Kosyakov:2020wxv, Russo:2024xnh}. By constraining the ModMax parameter using astrophysical data, we effectively probe the validity of Maxwell's theory in the extreme environments surrounding BHs, where electromagnetic fields are expected to be strongest \cite{Flores-Alfonso:2020euz, Bandos:2020jsw, Barrientos:2022bzm}. 

This paper is organized as follows. In Sec.~\ref{Sec:metric}, we review the Einstein-ModMax theory and present the static, spherically symmetric BH solution and its associated spacetime metric. Section~\ref{Sec:V} is dedicated to the parameter estimation of the ModMax BH parameters. We use the EHT observations of M87* and Sgr A*, the GRAVITY observation for Sgr A*, and a combined analysis to place tighter constraints on the parameters of ModMax BH. In Sec.~\ref{AccrDisk}, we investigate the physical properties of optically thin accretion disks in the ModMax BH spacetime, calculating and numerically analyzing the energy flux, temperature distribution, and emission spectra. We also discuss the impact of the ModMax BH parameters on these observables. Finally, we present the conclusions in Sec.~\ref{Sec:conclusion}.

\section{The spacetime metric and the dynamics of massless particles}\label{Sec:metric}

The ModMax theory is a nonlinear extension of Maxwell electrodynamics that preserves both conformal invariance and electromagnetic duality invariance. Its Lagrangian density can be written as \cite{Kosyakov:2020wxv,Bandos:2020hgy}
\begin{eqnarray} 
{\cal L}_{\rm MM}  =  - \frac{1}{2}\left( {s\cosh \nu - \sqrt {s^2  + p^2 } \sinh \nu} \right)\, ,
\end{eqnarray}
where $\nu$ is a dimensionless parameter and describes the nonlinear electrodynamics parameter of the ModMax theory. The two Lorentz-invariants, $s$ and $p$, are defined in terms of the electromagnetic field-strength tensor as
\begin{eqnarray} \label{eq.sp}
s = \frac{1}{2}F_{\mu \nu } F^{\mu \nu} \qquad \mbox{and} \qquad p = \frac{1}{2}F_{\mu \nu } \tilde F^{\mu \nu } \, .
\end{eqnarray} 
It is worth noting that these quantities, $s$ and $p$, are the two independent electromagnetic Lorentz invariants in four dimensions. In particular, the former measures the difference between the magnetic and electric field contributions, while the latter, $p\sim\mathbf{E}\cdot\mathbf{B}$, characterizes their relative orientation. The electromagnetic field-strength tensor $F_{\mu\nu}$ is obtained from the electromagnetic four-potential $A_{\mu}$ according to
\begin{eqnarray}
F_{\mu\nu}
=\partial_{\mu}A_{\nu}
\partial_{\nu}A_{\mu}\, ,
\end{eqnarray}
while dual tensor is defined by 
\begin{eqnarray}
\tilde{F}_{\mu\nu}
=\frac{1}{2}
\varepsilon_{\mu\nu\alpha\beta}
F^{\alpha\beta}\, ,
\end{eqnarray}
where $\varepsilon_{0123}=\sqrt{-g}$ with $g=\det(g_{\mu\nu})$ being the determinant of the spacetime metric. Equivalently, introducing the electromagnetic two-form, 
$ {\bf F} =\frac{1}{2}F_{\mu\nu}\,dx^{\mu}\wedge dx^{\nu}$,
its dual can be expressed in differential-form notation as ${\bf \tilde F}=\star{\bf F}$, where $\star$ denotes the Hodge dual operator associated with the spacetime metric. An important aspect of ModMax electrodynamics is that the parameter $\nu$ governs the deviation from standard Maxwell electrodynamics. In the Maxwell limit $\nu\rightarrow0$, one has
\begin{eqnarray}
{\cal L}_{\rm MM}\Big|_{\nu=0}=-\frac{s}{2}
-\frac{1}{4}F_{\mu\nu}F^{\mu\nu}\, ,
\end{eqnarray}
which is precisely the standard Maxwell Lagrangian density. Hence, nonzero values of the parameter $\nu$ encode nonlinear corrections to conventional electrodynamics. For small values of the ModMax parameter, ($\nu\ll1$), the Lagrangian may be expanded as
\begin{eqnarray}
{\cal L}_{\rm MM}
\simeq -\frac{s}{2} + \frac{\nu}{2}\sqrt{s^{2}+p^{2}}
-\frac{\nu^{2}}{4}s + {\cal O}(\nu^{3})\, ,
\end{eqnarray}
which explicitly shows how the leading nonlinear correction is governed by the combination $\sqrt{s^{2}+p^{2}}$. Therefore, even relatively small values of $\nu$ can modify electromagnetic configurations when the field invariants become sufficiently large, particularly in strong field environments. Despite its nonlinear nature, ModMax electrodynamics maintains conformal invariance and continuous electric-magnetic duality symmetry, while reducing to Maxwell theory when $\nu=0$. Causality and consistent propagation of electromagnetic perturbations require $\nu \geq 0$, which corresponds exactly to Maxwell electrodynamics \cite{Sorokin:2021tge}. Therefore, $\nu$ quantifies nonlinear deviations from Maxwell electrodynamics, which are generally expected to be small in physically viable regimes.

The nonlinear electromagnetic field in ModMax theory can be conveniently described by introducing the two-form associated with the "material" or constitutive field strength \cite{BallonBordo:2020jtw,Barrientos:2022bzm},
\begin{eqnarray}
\label{eq.TensorMaterial}
{\bf E}=2\left(
f_s {\bf F}
+
f_p {\bf \tilde F}
\right),
\end{eqnarray}
where ({\bf F}) is the electromagnetic field-strength two-form and ${\bf \tilde F}=\star{\bf F}$ denotes its Hodge dual. The quantities $f_s$ and $f_p$ characterize the response of the nonlinear electromagnetic theory to the two electromagnetic invariants and are defined by
\begin{eqnarray}
f_s=\frac{\partial {\cal L}_{\rm MM}}{\partial s},
\qquad
f_p =\frac{\partial {\cal L}_{\rm MM}}{\partial p}.
\end{eqnarray}
Taking into account the ModMax Lagrangian, the constitutive field strength can be written as 
\begin{eqnarray}
{\bf E}&=&-\left(\cosh\nu -\frac{s}{\sqrt{s^2+p^2}}\sinh\nu
\right){\bf F}\nonumber\\ &&+ \frac{p}{\sqrt{s^2+p^2}}
\sinh\nu\,{\bf \tilde F}\, .
\end{eqnarray}
Note that the above equation represents a nonlinear relation between the electromagnetic field $\bf{F}$ and the field $\bf{E}$. In contrast to Maxwell electrodynamics, where the corresponding relation is linear, the coefficients appearing here depend explicitly on the electromagnetic invariants $s$ and $p$.

The nonlinear electromagnetic field acts as a source of spacetime curvature through its stress-energy tensor. The electromagnetic stress-energy tensor of ModMax electrodynamics can be written as
\begin{eqnarray}
\label{eq.Tmn}
T_{\mu\nu}=\frac{1}{4\pi} \left(
s\,g_{\mu\nu}-2F_{\mu\kappa}F_{\nu\lambda}g^{\kappa\lambda}
\right)f_s\, .
\end{eqnarray}
It is worth noting that conformal invariance is an important nature of ModMax theory, i.e., its electromagnetic stress-energy tensor is traceless, $T^{\mu}_{\mu}=0$, as in ordinary Maxwell electrodynamics. Also, the electromagnetic field {\bf E} provides a natural definition of the conserved electric charge. For a closed and spacelike two-dimensional surface $\Sigma$, the electric charge enclosed by $\Sigma$ is given by
\begin{eqnarray}
\label{eq.Qe}
Q_e=\frac{1}{4\pi} \int_{\Sigma} \star{\bf E}\, ,
\end{eqnarray}
which is the flux of the constitutive field, rather than simply that of the Maxwell field strength, that determines the conserved electric charge.
\begin{figure*} [htb!]
    \centering
    \includegraphics[scale=0.35]{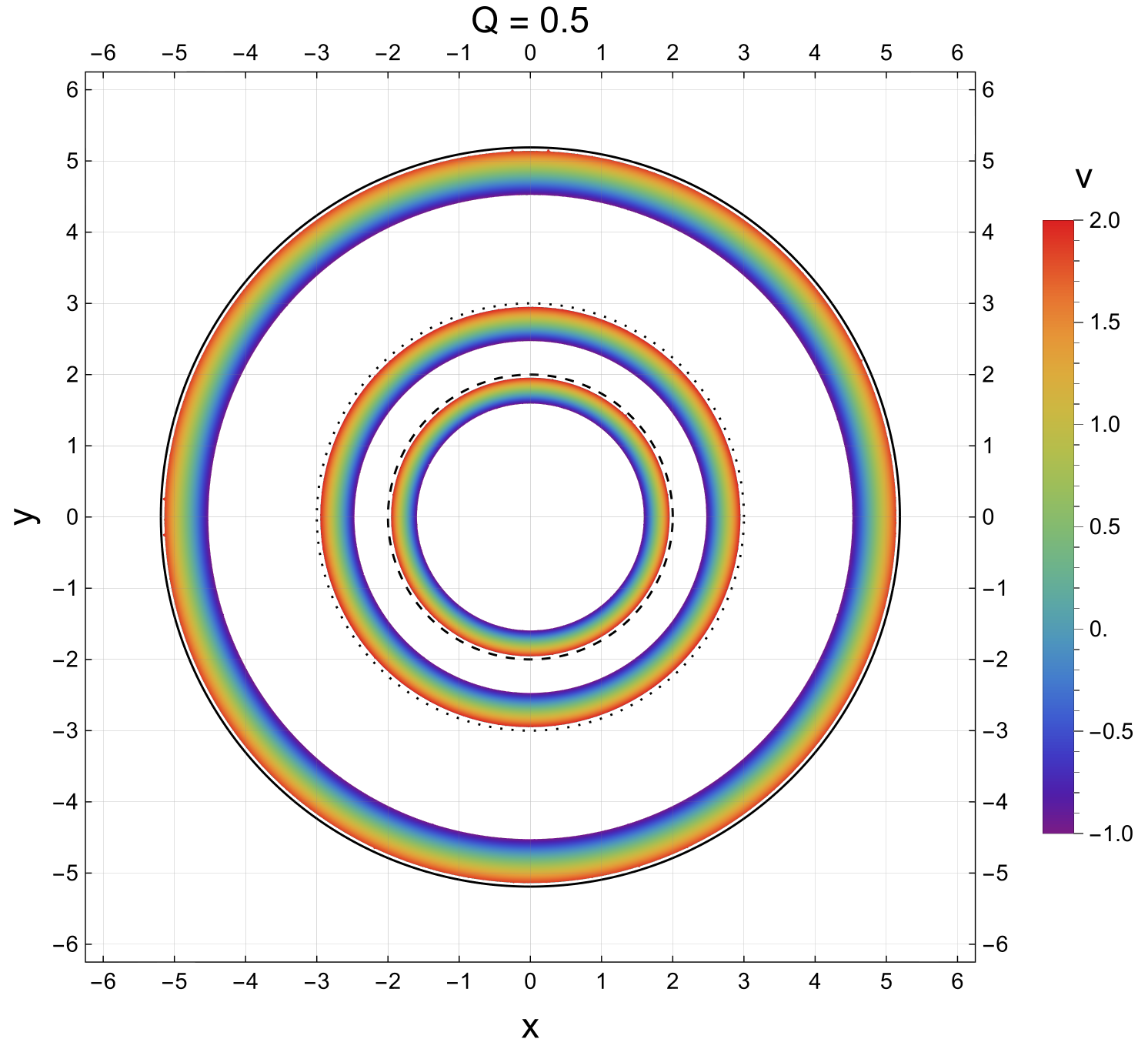}
    \includegraphics[scale=0.35]{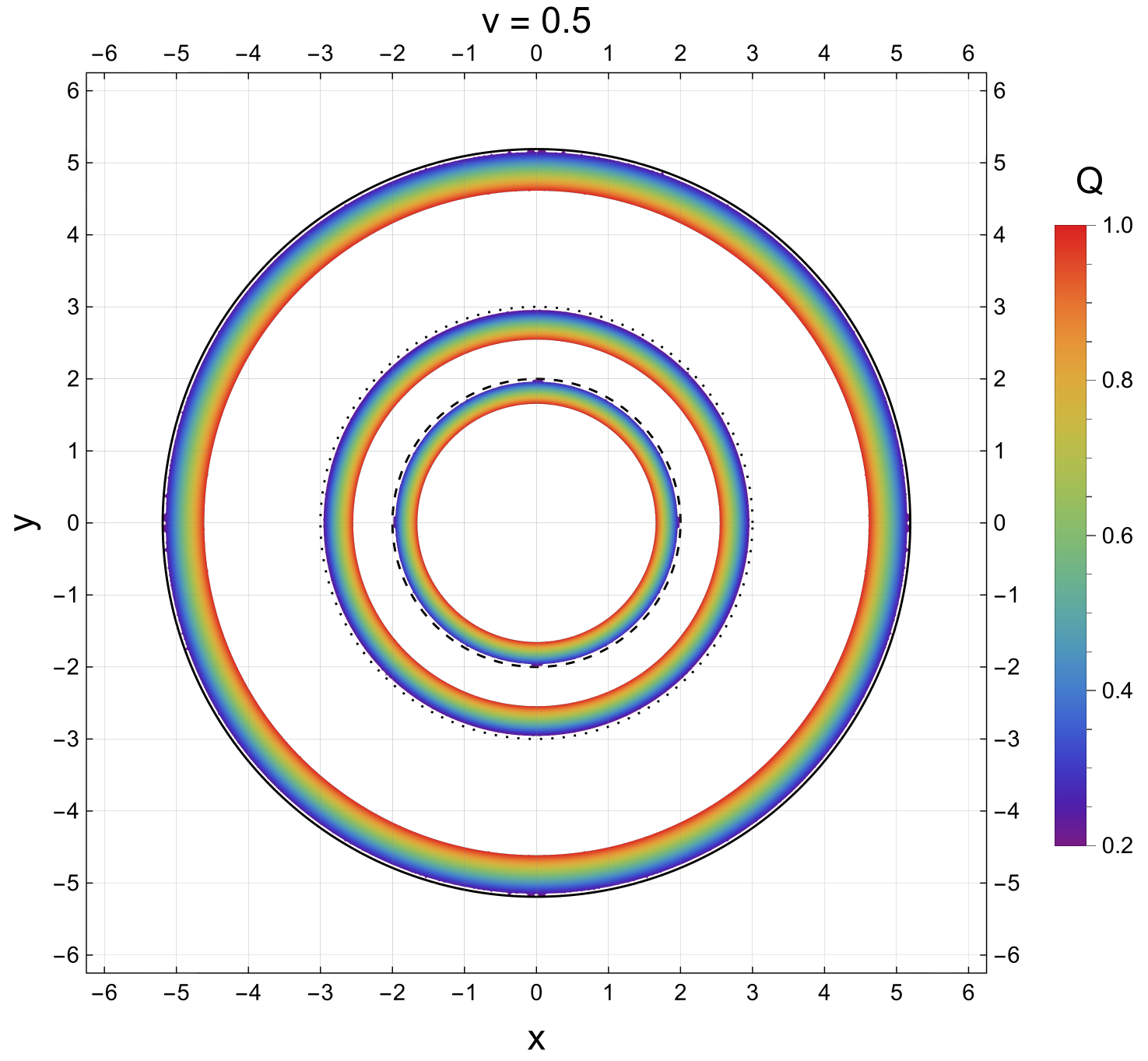}
    
    \caption{%
        \textit{Left panel:}  Variation of the event-horizon, photon-sphere, and shadow radii with the nonlinearity parameter $v$ for fixed $Q=0.5$. \textit{Right panel:} Variation of the event-horizon, photon-sphere, and shadow radii with the total charge $Q$ for fixed $v=0.5$. The innermost, middle, and outermost colored bands correspond to the event-horizon, photon-sphere, and shadow radii, respectively, whereas the dashed, dotted, and solid black curves denote the corresponding Schwarzschild values.
    }
 \label{fig:shadow}
\end{figure*}
\begin{figure*} 
    \centering
    \includegraphics[width = \textwidth]{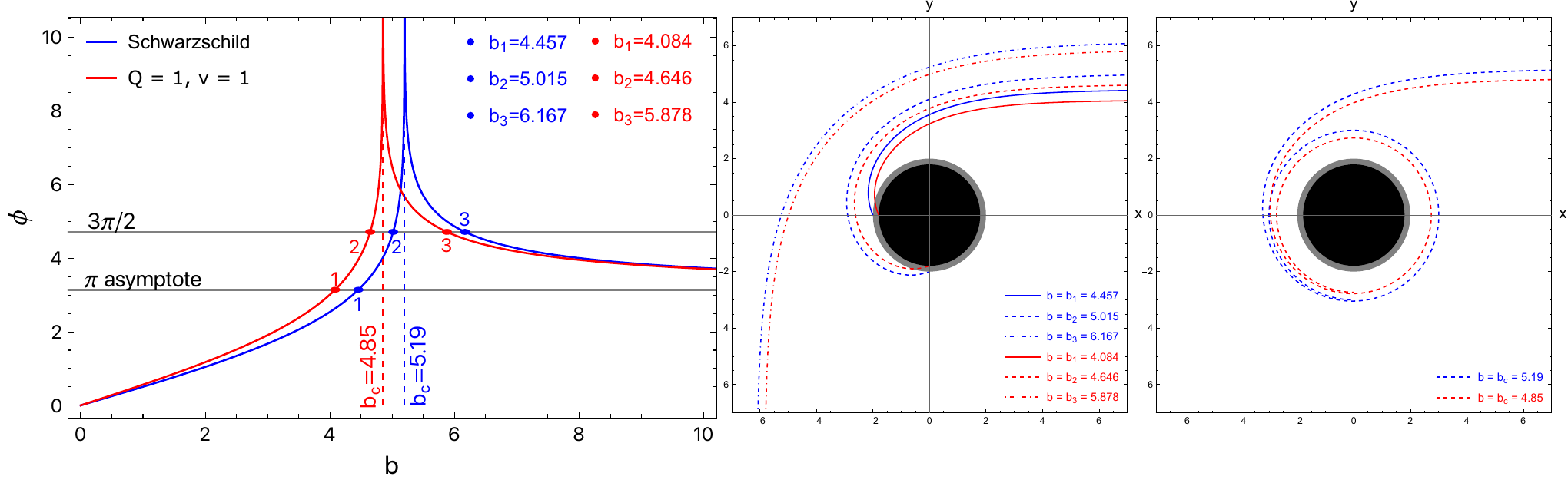}

    \caption{
        \textit{Left panel:} Total change in the azimuthal angle $\phi$ as a
        function of the impact parameter $b$ for the Schwarzschild BH and
        the ModMax BH with $Q=1$ and $v=1$. The marked points $b_1$
        correspond to $\phi=\pi$, while $b_2$ and $b_3$ correspond to
        $\phi=3\pi/2$. The vertical dashed lines mark the critical impact
        parameters $b_c=5.19$ and $b_c=4.85$ for the Schwarzschild and ModMax black
        holes, respectively.
        \textit{Middle panel:} Photon trajectories for the corresponding values of
        $b_1$, $b_2$, and $b_3$.
        \textit{Right panel:} Critical photon trajectories with $b=b_c$. 
    }

 \label{fig:deflection}
\end{figure*}

To incorporate the gravitational backreaction of the ModMax electromagnetic field, we consider the four-dimensional Einstein--ModMax action \cite{Flores-Alfonso:2020euz,BallonBordo:2020jtw,Barrientos:2022bzm},
\begin{eqnarray}
S=\frac{1}{16\pi} \int d^4x\sqrt{-g} \left(
R-4{\cal L}_{\rm MM} \right)\, ,
\end{eqnarray}
where $R$ is the Ricci scalar associated with the spacetime metric $g_{\mu\nu}$. The first term corresponds to the Einstein--Hilbert gravitational action, whereas the second accounts for the ModMax electromagnetic field.

Variation of this action with respect to the metric tensor $g_{\mu\nu}$ yields the Einstein field equations
\begin{eqnarray}
\label{eq.EinsteinEq}
R_{\mu\nu}
-\frac{1}{2}g_{\mu\nu}\,R=8\pi T_{\mu\nu},
\end{eqnarray}
where $T_{\mu\nu}$ is the ModMax electromagnetic stress-energy tensor given in Eq.~(\ref{eq.Tmn}). Hence, the nonlinear electromagnetic field determines the matter source responsible for modifying the spacetime geometry. On the other hand, variation with respect to the electromagnetic four-potential $A_\mu$ leads to the generalized source-free electromagnetic equation
\begin{eqnarray}
\label{eq.source-free}
\nabla^\mu E_{\mu\nu}=0\, ,
\end{eqnarray}
where $E_{\mu\nu}$ denotes the components of the constitutive two-form defined in Eq.~(\ref{eq.TensorMaterial}). In differential-form notation, this electromagnetic field equation takes the form as $d\star{\bf E}=0$, with the Bianchi identity $\nabla_{[\lambda}F_{\mu\nu]}=0$ or, equivalently, $d{\bf F}=0$.
Therefore, the complete electromagnetic part is governed by the following pair
\begin{eqnarray}
d{\bf F}=0,
\qquad
d\star{\bf E}=0\, .
\end{eqnarray}
In the linear limit $\nu\rightarrow0$, one can obtain 
\begin{eqnarray}
f_s\rightarrow-\frac{1}{2} \qquad \mbox{and} \qquad f_p\rightarrow0\, ,
\end{eqnarray}
for which the constitutive relation for ModMax
electrodynamics reduces to the ordinary Maxwell field strength. As a result, the generalized field equation reduces to
\begin{eqnarray}
\nabla_\mu F^{\mu\nu}=0\, ,
\end{eqnarray}
thereby recovering the standard source-free Maxwell equations. The Einstein--ModMax system therefore reduces to the Einstein--Maxwell theory as the nonlinear ModMax parameter $\nu$ approaches zero.

The spacetime geometry of the Einstein--ModMax BH is described by the following line element \cite{Flores-Alfonso:2020euz,ModMax2025}:
\begin{equation} \label{spacetime}
ds^2 = -f(r) dt^2 + f(r)^{-1} dr^2 + r^2 ( d\theta^2 + \sin^2 \theta d\phi^2),
\end{equation}
where 
\begin{equation} 
f(r) = 1 - \frac{2M}{r} + \frac{e^{-v}Q^2}{r^2}\, .
\label{metric_funct}
\end{equation}
Here, $M$ is the BH mass and $v$ represents the nonlinearity parameter, acting as a charge-screening factor that modifies the effective charge of BHs, while $Q^{2}=Q_{e}^{2}+Q_{m}^{2}$ is the total dyonic charge, with $Q_{e}$ and $Q_{m}$ denoting the electric and magnetic charges, respectively. For further analysis, we shall restrict our attention to the total charge $Q$. 

The Lagrangian that governs the motion of a massless particle, such as a photon, in the spacetime of the ModMax BH can be written as~\cite{1983mtbh.book.....C}
\begin{equation}
\mathcal{L}
=\frac{1}{2}g_{\mu\nu}
\frac{dx^{\mu}}{d\lambda}
\frac{dx^{\nu}}{d\lambda},
\end{equation}
where $\lambda$ denotes an affine parameter along the particle trajectory. The four-momentum of a photon can be expressed as follows:
\begin{equation}\label{momentum-photon}
    p^{\alpha}=\frac{dx^{\alpha}}{d\lambda},
\end{equation}
Since the spacetime metric is independent of the coordinates $t$ and $\phi$,
the photon energy $E$ and angular momentum $L$ are conserved and can be expressed as~\cite{Misner73}
\begin{equation}\label{conservation}
    p_t = g_{tt} \frac{dt}{d\lambda}=-E,\qquad p_\phi=g_{\phi\phi}\frac{d\phi}{d\lambda}=L,
\end{equation}

Using Eq.~\eqref{conservation} together with the normalization
condition for null geodesics,
$g_{\mu\nu}\dot{x}^{\mu}\dot{x}^{\nu}=0$, the equations governing photon propagation can be written as follows:
\begin{equation}
    \frac{dt}{d\lambda}=\frac{E}{f(r)},
\end{equation}
\begin{equation}
    \frac{d\phi}{d\lambda}=\frac{L}{r^2},
\end{equation}    
\begin{equation}
    \left(\frac{dr}{d\lambda}\right)^2=E^2-f(r)\frac{L^2}{r^2}.
\end{equation}

Introducing the rescaled affine parameter $\lambda'=L\lambda$, 
the equations governing light propagation can be rewritten as
\begin{equation}
    \frac{dt}{d\lambda'}=\frac{1}{b f(r)},
\end{equation}
\begin{equation}
    \frac{d\phi}{d\lambda'}=\frac{1}{r^2},
\end{equation}    
\begin{equation}
    \left(\frac{dr}{d\lambda'}\right)^2
    =\frac{1}{b^2}-\frac{f(r)}{r^2},
\end{equation}
where $b=L/E$ denotes the impact parameter of the photon. Using the above equations, we obtain the radial equation in terms of the
azimuthal angle $\phi$ as
\begin{equation} \label{radial-photon-equation}
    \left(\frac{dr}{d\phi}\right)^2
    =r^4\left(\frac{1}{b^2}-\frac{f(r)}{r^2}\right)
    \equiv V_{\rm eff},
\end{equation}
where $V_{\rm eff}$ denotes the effective potential for photon motion. The photon-sphere radius $r_{\rm ph}$ is determined from the condition
\begin{equation}
\left.\frac{dV_{\rm eff}}{dr}\right|_{r=r_{\rm ph}}=0.
\end{equation}
For a distant observer, the apparent radius of the BH shadow is
determined by the critical impact parameter of photon trajectories,
$R_{\rm sh}=b_c$. Photons with $b=b_c$ asymptotically approach the photon
sphere and form the boundary of the shadow. The critical impact parameter
is obtained from the condition $V_{\rm eff}(r_{\rm ph})=0$ as
\begin{equation}
R_{\rm sh} = b_c=\frac{r_{\rm ph}}{\sqrt{f(r_{\rm ph})}}.
\end{equation}

Fig.~\ref{fig:shadow} shows the dependence of the event-horizon, photon-sphere, and shadow radii on the ModMax BH parameters $Q$ and $v$, while the dashed, dotted, and solid black curves indicate their Schwarzschild values. 
For fixed $Q=0.5$, increasing $v$ shifts all three characteristic radii toward their Schwarzschild values. For fixed $v=0.5$, the Schwarzschild limit is recovered at $Q=0$, while increasing $Q$ monotonically decreases the event-horizon, photon-sphere, and shadow radii. Thus, the total charge reduces the apparent shadow size, whereas larger values of $v$ weaken the effect of the charge.

By introducing the new variable $u=1/r$, the radial equation
for photon motion given in Eq.~\eqref{radial-photon-equation} can be expressed as
\begin{equation} \label{eq:u-equation}
    \left( \frac{du}{d\phi} \right)^2 = \frac{1}{b^2} - u^2 f\left( \frac{1}{u} \right)\equiv G(u).
\end{equation}
\renewcommand{\arraystretch}{1.5}
\begin{table}[t]
\centering
\resizebox{.5\textwidth}{!}{
\begin{tabular}{lcccc}
\hline
Object & $\theta_{\rm sh}$ ($\mu$as) & Distance $D$ & Mass $M$ & Data \\
\hline
M87$^{\star}$ & $42 \pm 3$ & $16.8 \pm 0.8$ Mpc & $(6.5 \pm 0.7)\times10^{9}\,M_{\odot}$ & EHT \\
Sgr A$^{\star}$ & $48.7 \pm 7$ & $8150 \pm 150$ pc & $(4.0^{+1.1}_{-0.6})\times10^{6}\,M_{\odot}$ & EHT \\
Sgr A$^{\star}$ & --- & $8275.9 \pm 8.6$ pc & $(4.299 \pm 0.012)\times10^{6} M_{\odot}$ & GRAVITY \\
\hline
\end{tabular}
}
\caption{Observational parameters of M87$^\star$ and Sgr~A$^\star$ obtained from the EHT and GRAVITY measurements (see details in \cite{EHT_2019ApJ875L6E,EHT2022ApJ930L12E,Gravity2024A&A692A242G}).}
\label{tab:shadow}
\end{table}
\begin{figure*}[htb!]
    \centering
\includegraphics[scale=0.35]{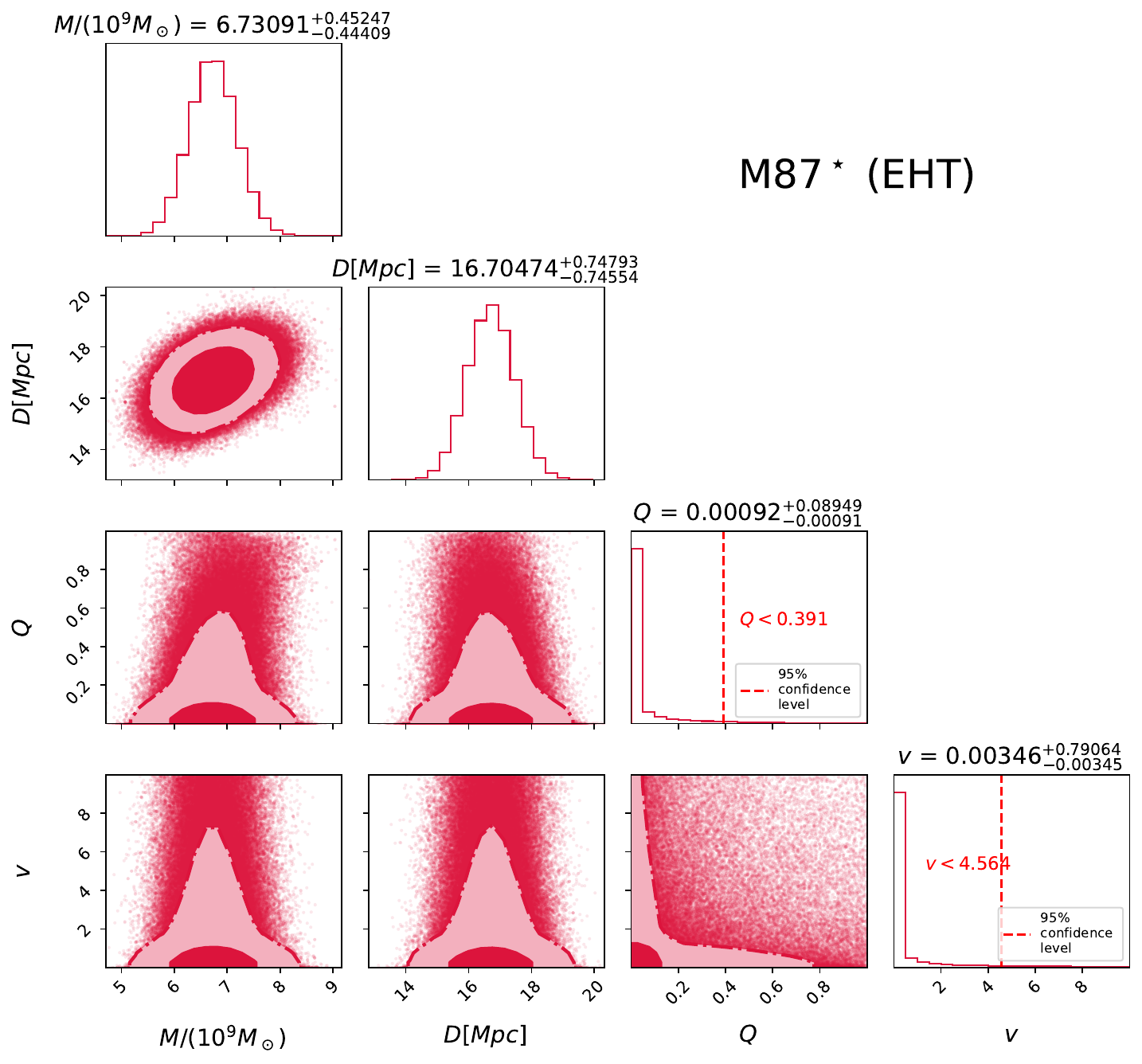}
\includegraphics[scale=0.35]{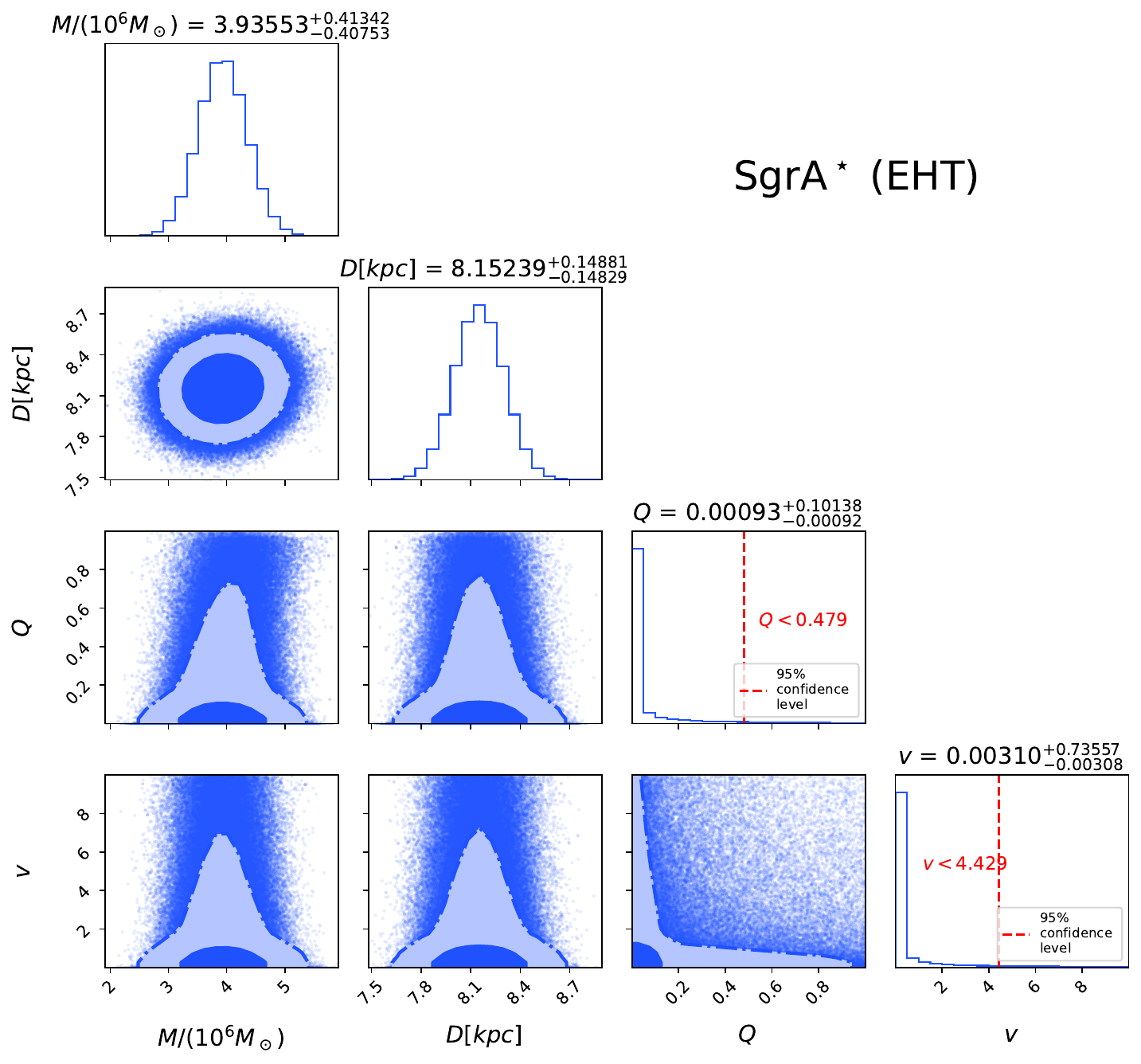}
\includegraphics[scale=0.35]{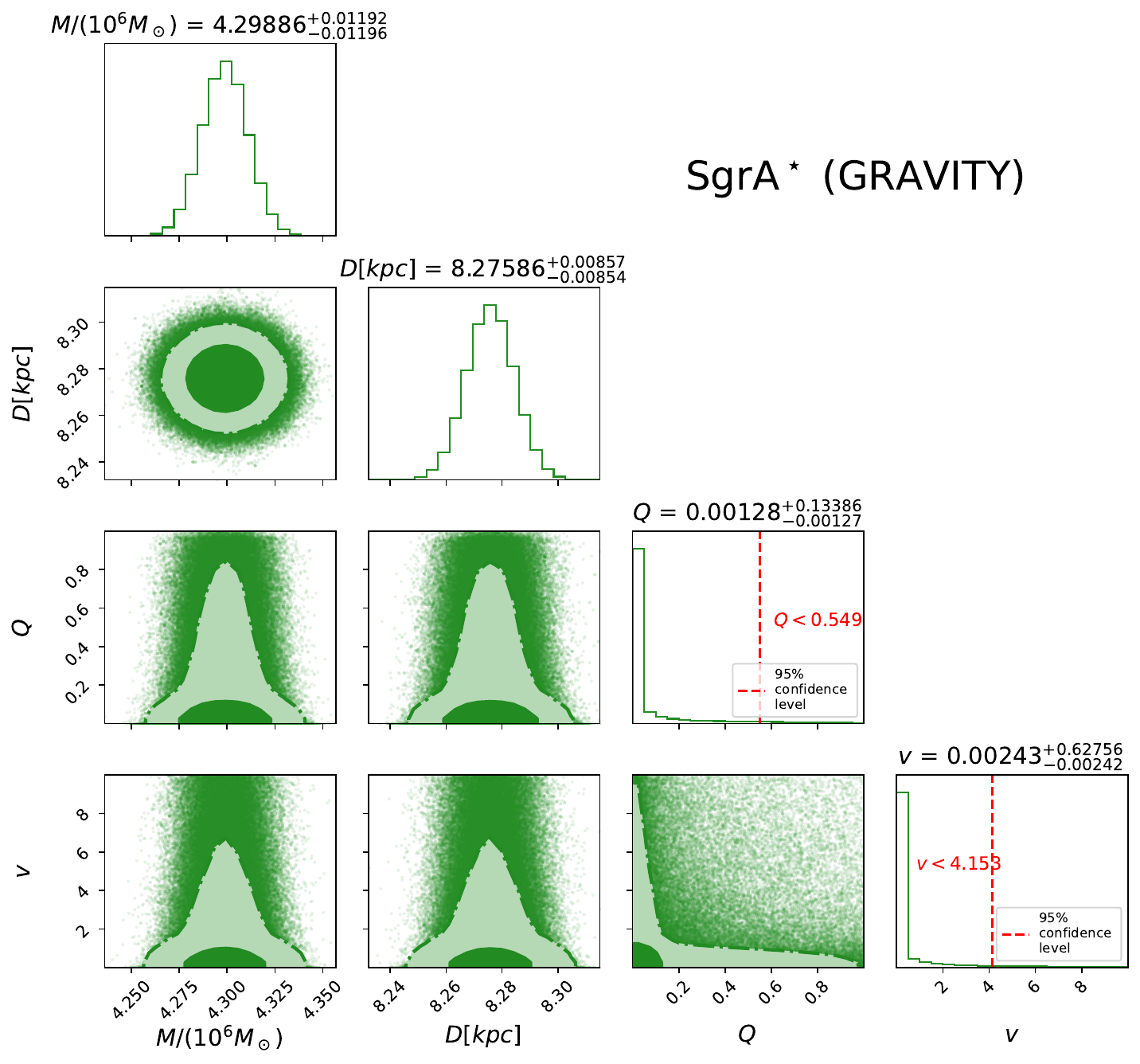}
\includegraphics[scale=0.35]{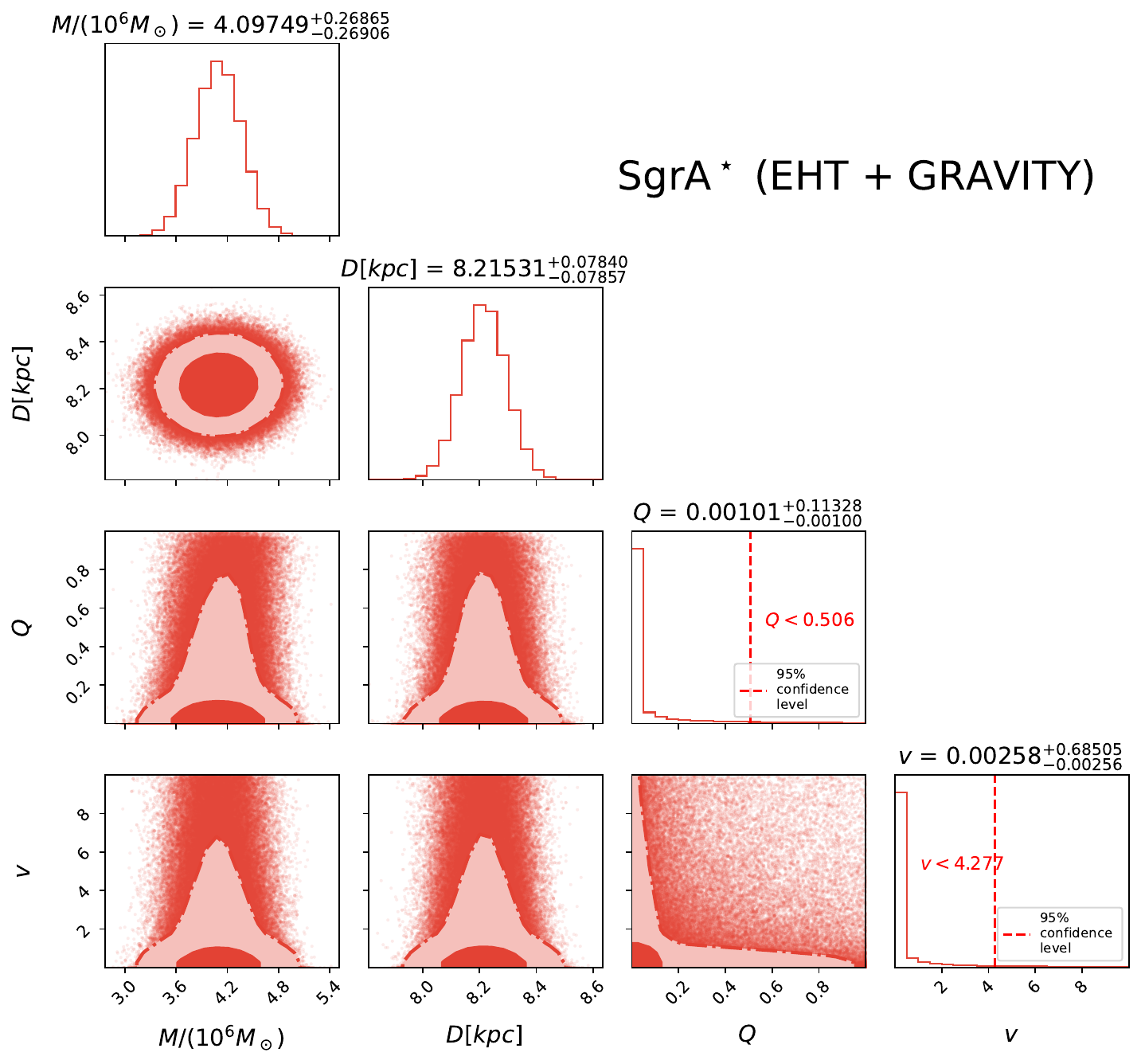}
    \caption{
        Posterior probability distributions of the model parameters $(M,D,Q,v)$ obtained from the shadow observations of M87$^\star$ and Sgr A$^\star$, where $Q$ is the total charge of the ModMax BH and $v$ is the nonlinearity parameter. The vertical red dashed lines show the 95\%  confidence-level upper limits on $Q$ and $v$.
    }
    \label{Fig.mcmc}
\end{figure*}

From Eq.~\eqref{eq:u-equation}, the photon trajectory depends on the value of
the impact parameter $b$ relative to its critical value $b_c$. For $b<b_c$,
a photon coming from infinity is captured by the BH. In this case,
we consider only the part of the trajectory outside the event horizon, and
the total change in the azimuthal angle is obtained by integrating from
spatial infinity, $u=0$, to the horizon, $u_h=1/r_h$, as
\begin{equation}
    \varphi = \int_{0}^{u_h}\frac{du}{\sqrt{G(u)}},
    \qquad b<b_c .
\end{equation}
Here, $r_h$ denotes the radius of the outer event horizon.

For $b>b_c$, the photon does not cross the event horizon. Instead, it reaches
a turning point at the closest approach, $r_{\min}$, and then escapes back to
infinity. Since the trajectory is symmetric with respect to the turning point,
the total change in the azimuthal angle is given by
\begin{equation}
    \varphi = 2\int_{0}^{u_{\rm min}}\frac{du}{\sqrt{G(u)}},
    \qquad b>b_c ,
\end{equation}
where $u_{\rm min}=1/r_{\min}$ is the minimum positive root of $G(u)=0$ corresponding to the turning point.

Fig.~\ref{fig:deflection} compares the total change in the azimuthal angle and the corresponding photon trajectories in the Schwarzschild and ModMax spacetimes. The azimuthal angle $\phi$ increases rapidly as the impact parameter approaches its critical value $b_{\rm c}$. For the ModMax BH with $Q=v=1$, the critical impact parameter is approximately $b_{\rm c}=4.85$, which is smaller than the Schwarzschild value $b_{\rm c}=5.19$. 
The middle and right panels further show that the ModMax geometry shifts the photon trajectories toward smaller radial distances relative to the Schwarzschild case. For $b=b_{\rm c}$, the photons asymptotically approach their respective unstable circular orbits at $r=r_{\rm ph}$.

\section{Parameter estimation for ModMax BH parameters}\label{Sec:V}
\begin{table*}[ht]
\centering
\begin{tabular}{lcccc}
\hline\hline
Parameter 
& M87$^\star$ (EHT) 
& Sgr A$^\star$ (EHT) 
& Sgr A$^\star$ (GRAVITY) 
& Sgr A$^\star$ (EHT+GRAVITY) \\
\hline

$M$ 
& $6.73091^{+0.45247}_{-0.44409}\times10^{9}\,M_\odot$
& $3.93553^{+0.41342}_{-0.40753}\times10^{6}\,M_\odot$
& $4.29886^{+0.01192}_{-0.01196}\times10^{6}\,M_\odot$
& $4.09749^{+0.26865}_{-0.26906}\times10^{6}\,M_\odot$ \\

$D$
& $16.70474^{+0.74793}_{-0.74554}\,\mathrm{Mpc}$
& $8.15239^{+0.14881}_{-0.14829}\,\mathrm{kpc}$
& $8.27586^{+0.00857}_{-0.00854}\,\mathrm{kpc}$
& $8.21531^{+0.07840}_{-0.07857}\,\mathrm{kpc}$ \\

$Q$
& $0.00092^{+0.08949}_{-0.00091}$
& $0.00093^{+0.10138}_{-0.00092}$
& $0.00128^{+0.13386}_{-0.00127}$
& $0.00101^{+0.11328}_{-0.00100}$ \\

$v$
& $0.00346^{+0.79064}_{-0.00345}$
& $0.00310^{+0.73557}_{-0.00308}$
& $0.00243^{+0.62756}_{-0.00242}$
& $0.00258^{+0.68505}_{-0.00256}$ \\

$Q$ (95\% C.L.)
& $Q<0.391$
& $Q<0.479$
& $Q<0.549$
& $Q<0.506$ \\

$v$ (95\% C.L.)
& $v<4.564$
& $v<4.429$
& $v<4.153$
& $v<4.277$ \\

\hline\hline
\end{tabular}
\caption{
Best-fit values and upper limits of the model parameters obtained
from the MCMC analysis of the M87$^\star$ and Sgr~A$^\star$ data. The
reported uncertainties correspond to the 68\% credible intervals, while
the upper limits on $Q$ and $v$ are given at the 95\% confidence level.
}
\label{tab:mcmc}
\end{table*}

The observed angular diameter of the BH shadow is given by
\begin{equation}
\theta_{\rm sh}=R_{\rm sh}\frac{2GM}{c^2D},
\end{equation}
where $R_{\rm sh}$ is the dimensionless shadow radius and $D$ denotes the distance from the observer to the BH. Based on the shadow measurements of Sgr A$^{\star}$ and M87$^{\star}$ given
in Table~\ref{tab:shadow}, we constrain the model parameters
$M$, $D$, $Q$, and $v$. The posterior distributions of these parameters are obtained through a Markov Chain Monte Carlo analysis implemented with
the Python package \textit{emcee}~\cite{Foreman_Mackey_2013}.

The likelihood function measures the agreement between the theoretical prediction of the angular shadow size, $\theta_{\rm sh}^{\rm theory}$, and the observed value, $\theta_{\rm sh}^{\rm obs}$. The corresponding log-likelihood is written as
\begin{equation}
    \log \mathcal{L}(M,D,Q,v)
    =
    -\frac{1}{2}
    \left[
    \frac{\theta_{\rm sh}^{\rm theory}(M,D,Q,v)
    -\theta_{\rm sh}^{\rm obs}}
    {\sigma_{\rm sh}^{\rm obs}}
    \right]^2 ,
\end{equation}
where $\sigma_{\rm sh}^{\rm obs}$ denotes the observational uncertainty in the shadow measurement.

For the BH mass $M$ and distance $D$, we adopt Gaussian priors
centered on their measured values, with standard deviations determined by
their observational uncertainties. Assuming that the two priors are independent,
the joint prior can be written as
\begin{equation}
\pi(M,D) \propto
\exp\left[
-\frac{1}{2}
\left(
\frac{M-M_0}{\sigma_M}
\right)^2
-\frac{1}{2}
\left(
\frac{D-D_0}{\sigma_D}
\right)^2
\right],
\end{equation}
where $M_0$ and $D_0$ are the measured values of the BH mass and
distance, while $\sigma_M$ and $\sigma_D$ denote their corresponding
observational uncertainties.

For the parameters $Q$ and $v$, we assume independent uniform priors: 
\begin{equation}
\pi(Q,v)= 
    \begin{cases} 
        1, & 0 \leq Q \leq 1,\; 0 \leq v \leq 10,\\ 
        0, & \text{otherwise}. 
    \end{cases} 
\end{equation}

Using Bayes' theorem, the posterior distribution of the model parameters
$\Theta=(M,D,Q,v)$ can be written as
\begin{equation}
P(\Theta|\mathcal{D}_{\rm obs})=
\frac{\mathcal{L}(\mathcal{D}_{\rm obs}|\Theta)\,\pi(\Theta)}
{\mathcal{Z}},
\end{equation}
where $\mathcal{D}_{\rm obs}$ represents the observational data,
$\pi(\Theta)$ is the joint prior, and $\mathcal{Z}$ is the normalization
factor.

Fig.~\ref{Fig.mcmc} shows the posterior probability distributions of the
model parameters obtained from the shadow constraints of M87$^{\star}$ and
Sgr~A$^{\star}$. The shaded contour regions represent the 68\% and 95\% confidence levels. The vertical dashed lines mark the
95\% confidence-level upper bounds on the ModMax BH parameters $Q$ and $v$.
The resulting best-fit values, together with the corresponding upper limits,
are listed in Table~\ref{tab:mcmc}.

From Tables~\ref{tab:shadow} and~\ref{tab:mcmc}, we find that the MCMC results for the BH mass and distance are consistent with the observational values listed in Table~\ref{tab:shadow}. The best-fit values of the ModMax BH parameters $Q$ and $v$ are close to zero, suggesting that the present shadow observations do not indicate significant charge or nonlinearity effects for M87$^{\star}$ and Sgr~A$^{\star}$. Instead, these observations mainly provide upper limits of these parameters. The strongest constraint on the total charge is obtained from M87$^\star$ (EHT), yielding $Q<0.391$ at the 95\% confidence level, while the tightest bound on the nonlinearity parameter is obtained from Sgr~A$^\star$ (GRAVITY), giving $v<4.153$.

\section{Images and physical properties of thin accretion disk around ModMax BH}\label{AccrDisk}

\begin{figure*} [htb!]
    \centering
    \includegraphics[scale=0.67]{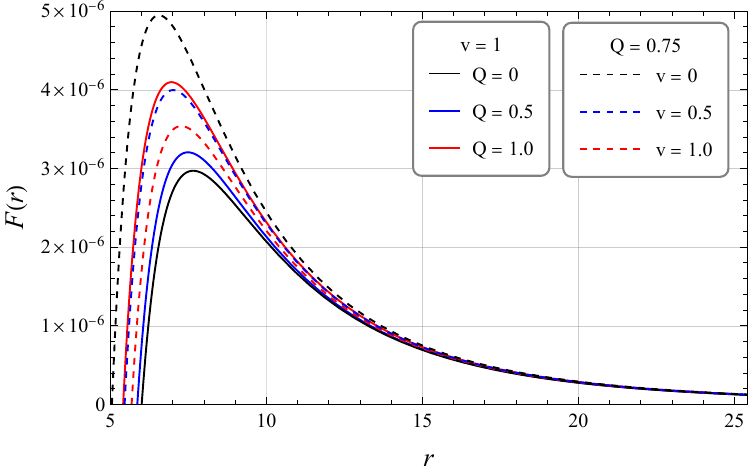}
    \includegraphics[scale=0.65]{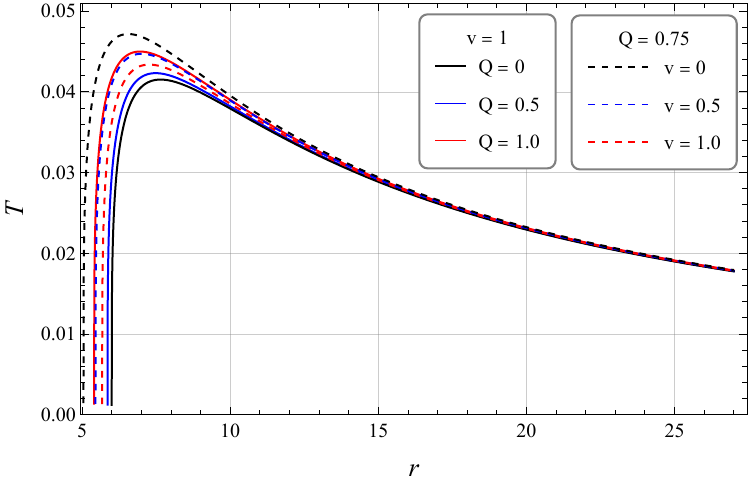}
    
    \caption{%
        Radial profiles of the radiative energy flux $F(r)$ (left panel) and temperature $T(r)$ (right panel) of a thin accretion disk around a ModMax BH. The solid curves correspond to a fixed nonlinearity parameter $v=1$ with $Q=0$, $Q = 0.5$, and $Q = 1.0$, whereas the dashed curves correspond to a fixed total charge $Q=0.75$ with $v=0$, $v = 0.5$, and $v = 1.0$.
    }
 \label{fig:fux&temp}
\end{figure*}

\begin{figure*}[htb!]
\begin{tabular}{ccc}
  \includegraphics[scale=0.35]{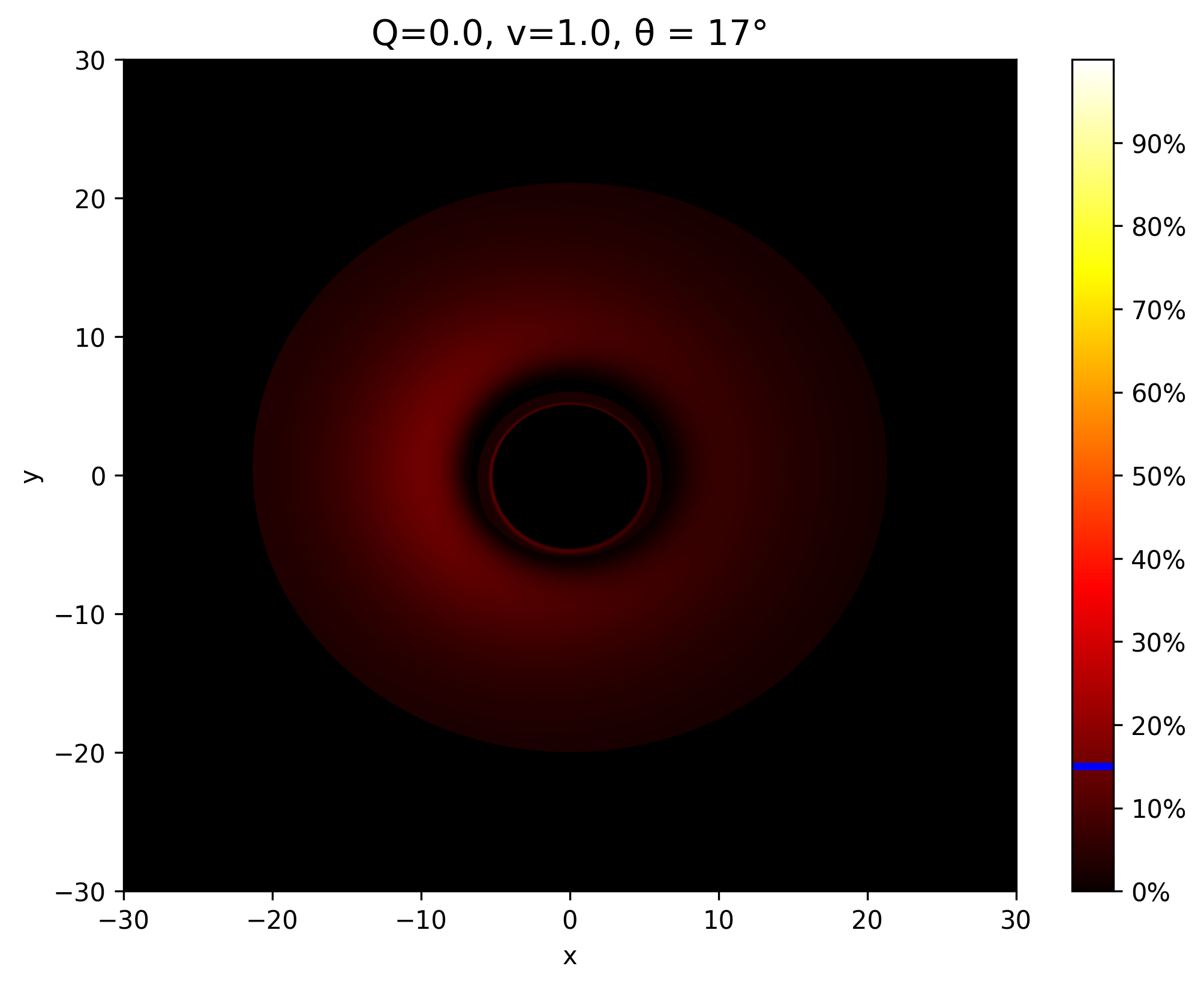}\hspace{-0.2cm}
  \includegraphics[scale=0.35]{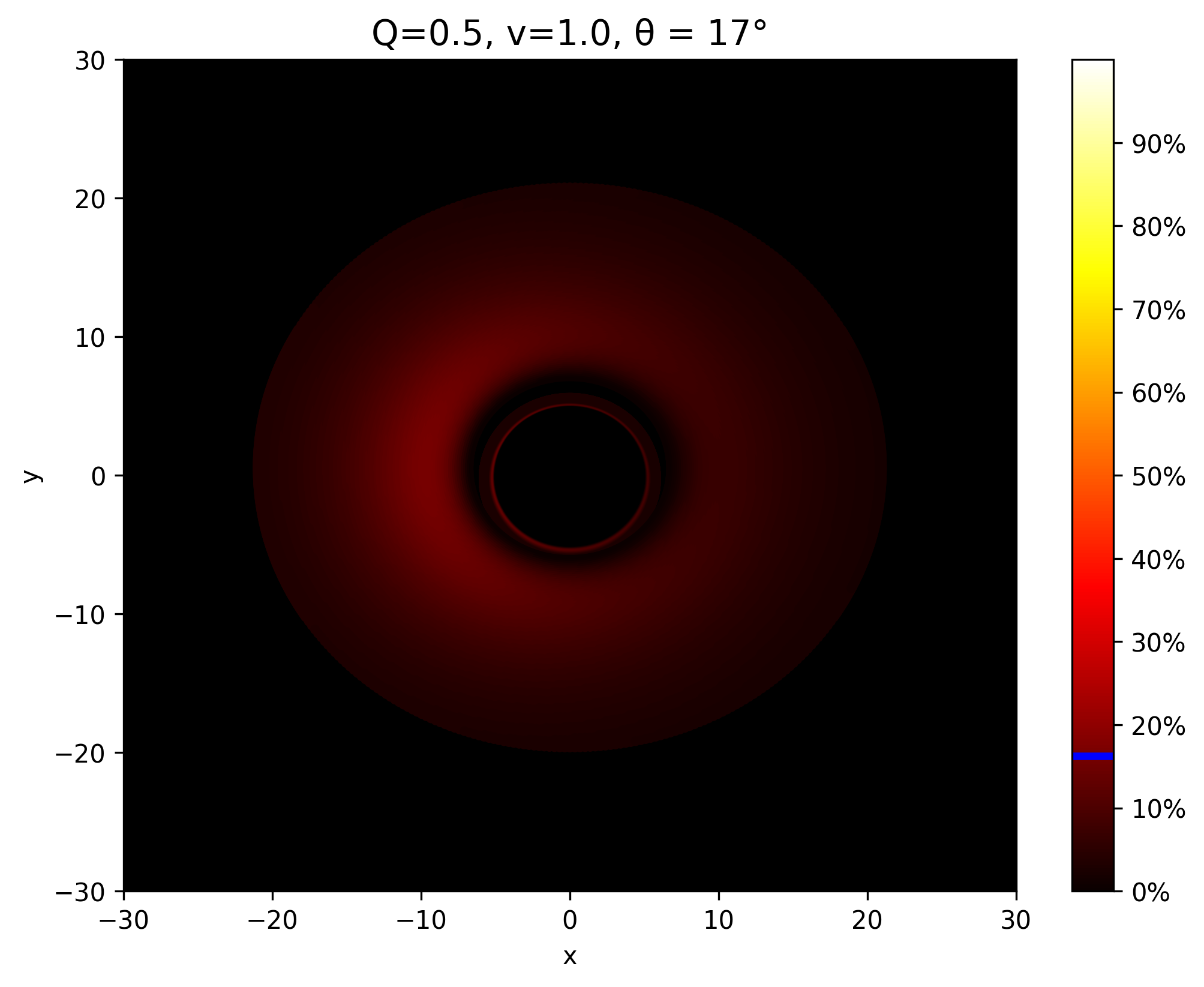}\hspace{-0.2cm}
  \includegraphics[scale=0.35]{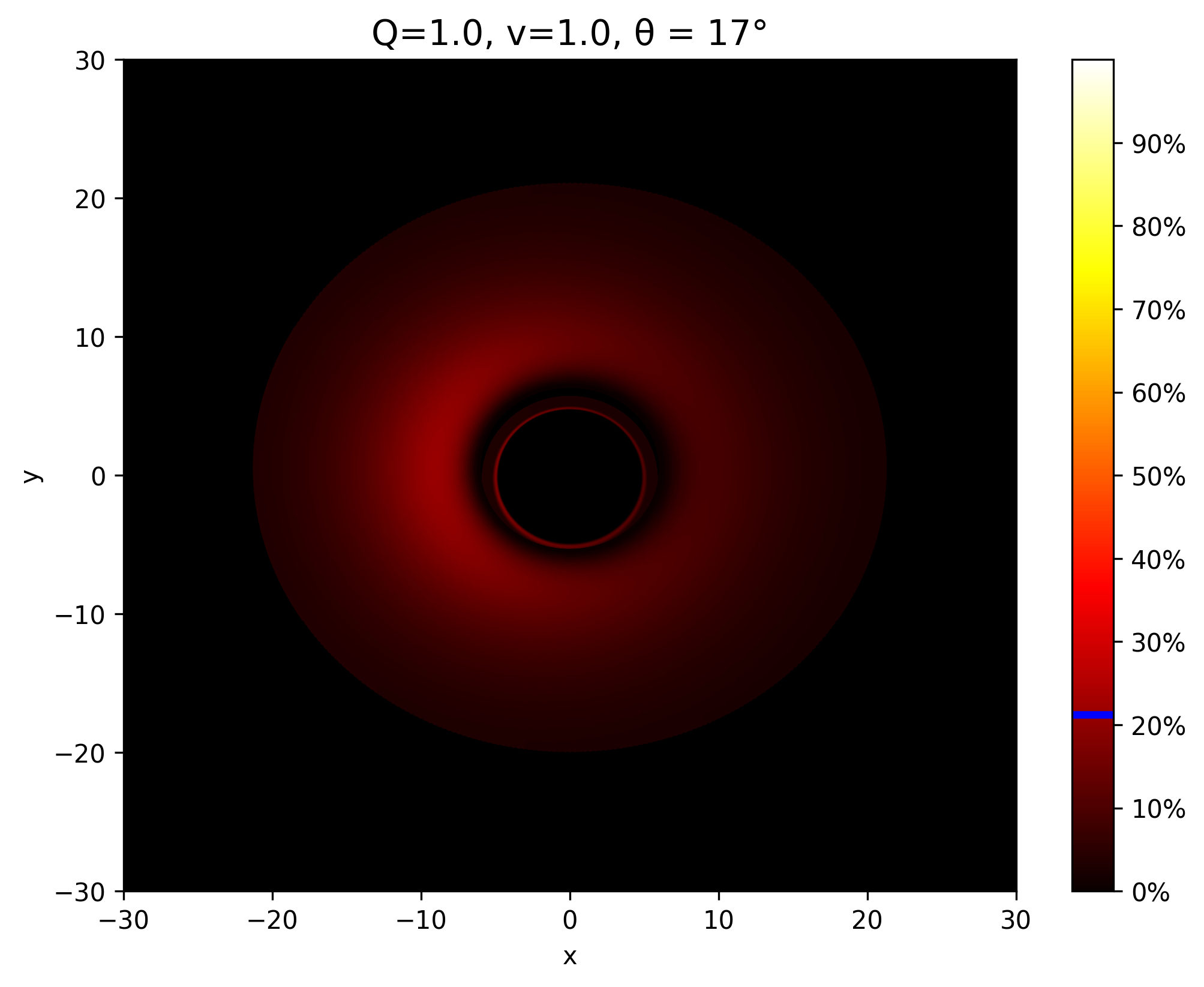}\\
  \includegraphics[scale=0.35]{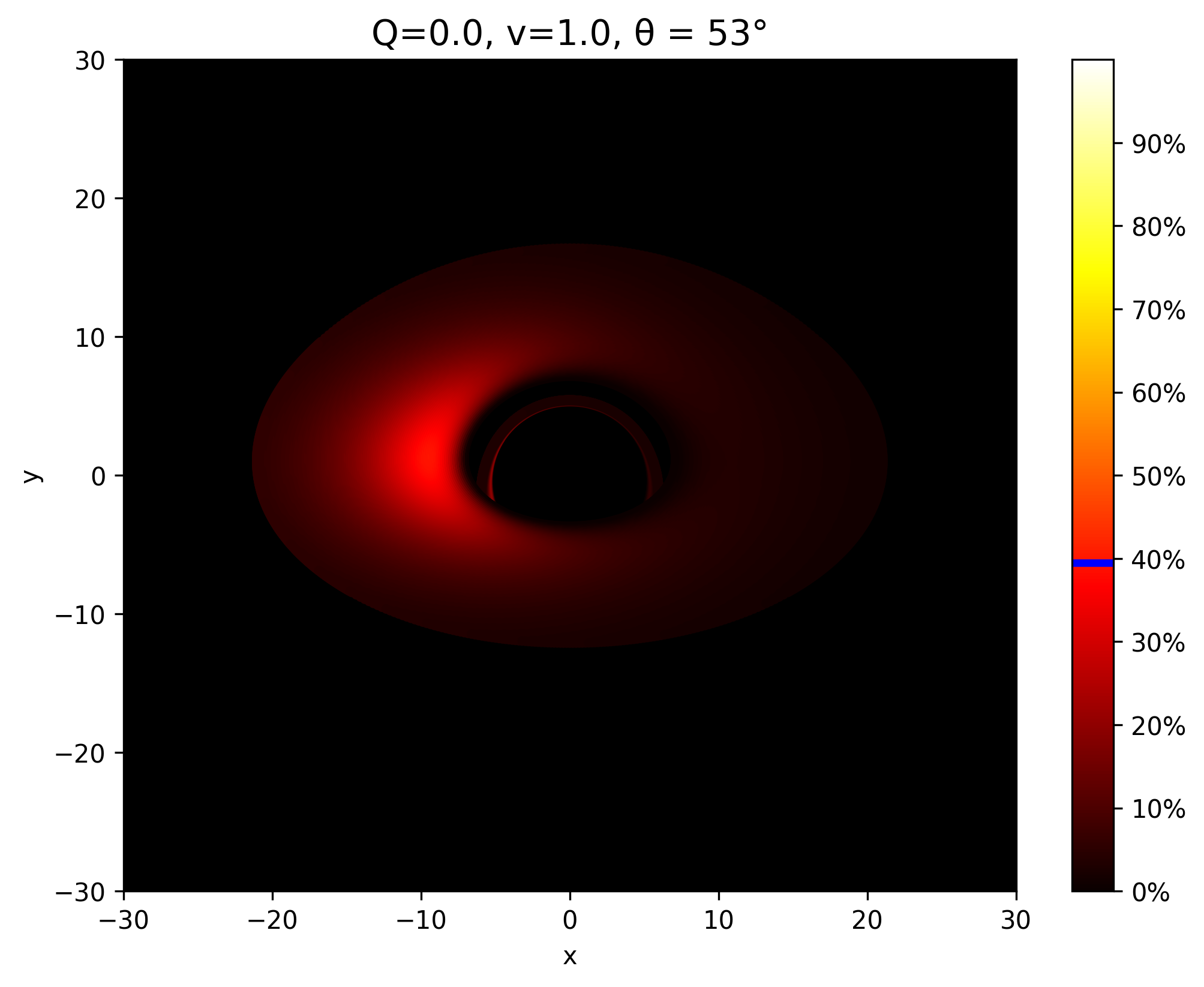}\hspace{-0.2cm}
  \includegraphics[scale=0.35]{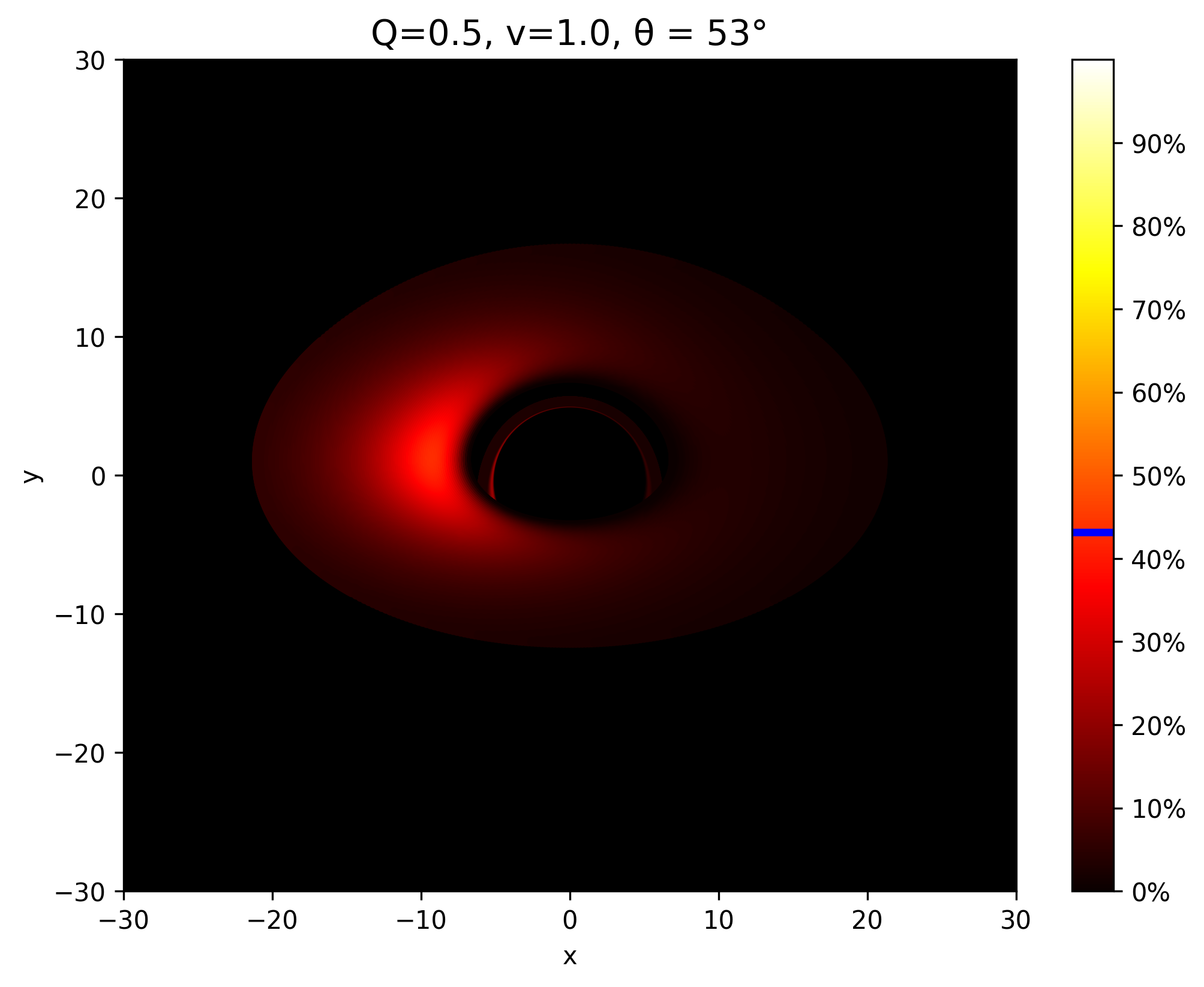}\hspace{-0.2cm}
  \includegraphics[scale=0.35]{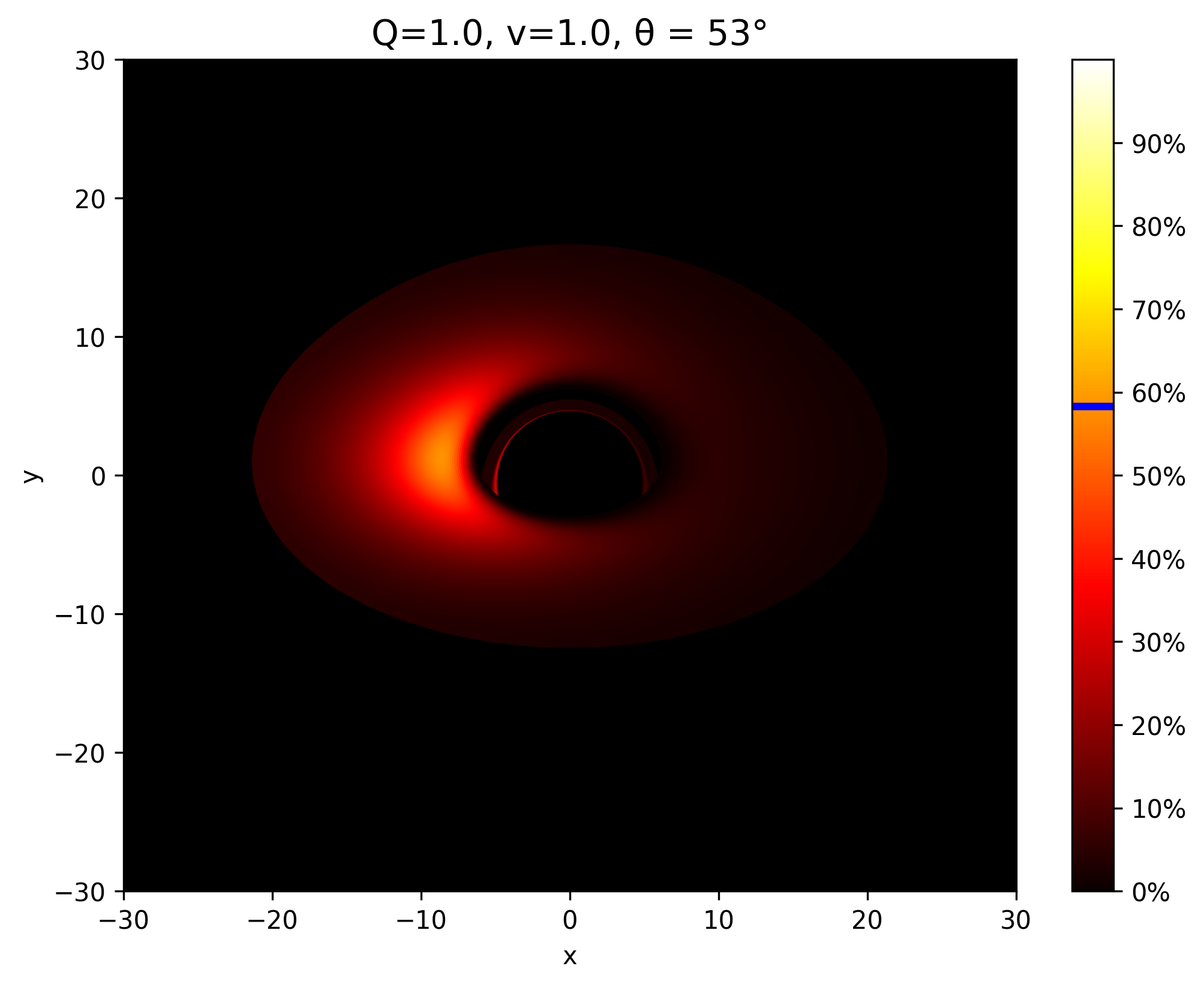}\\
  \includegraphics[scale=0.35]{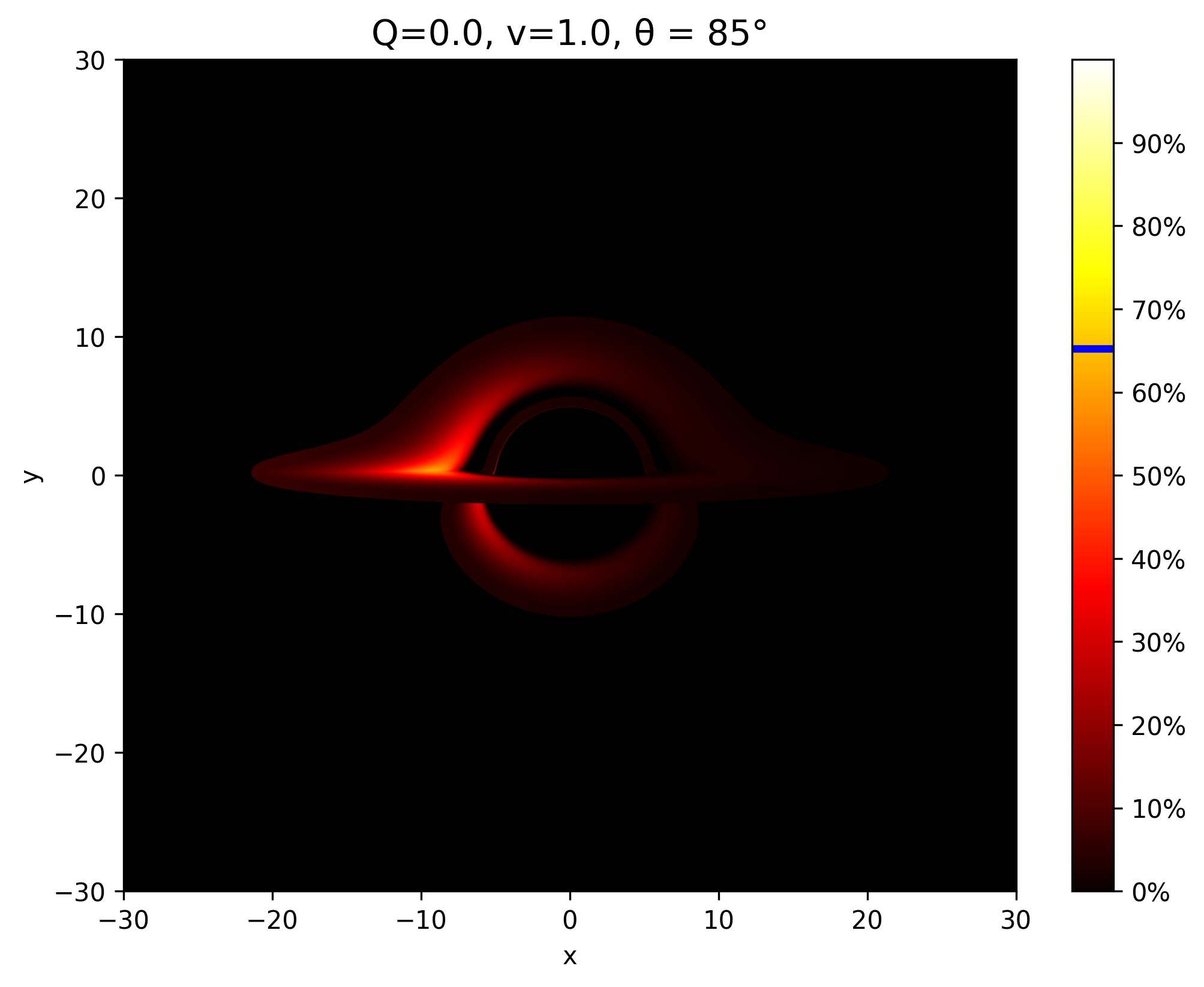}\hspace{-0.2cm}
  \includegraphics[scale=0.35]{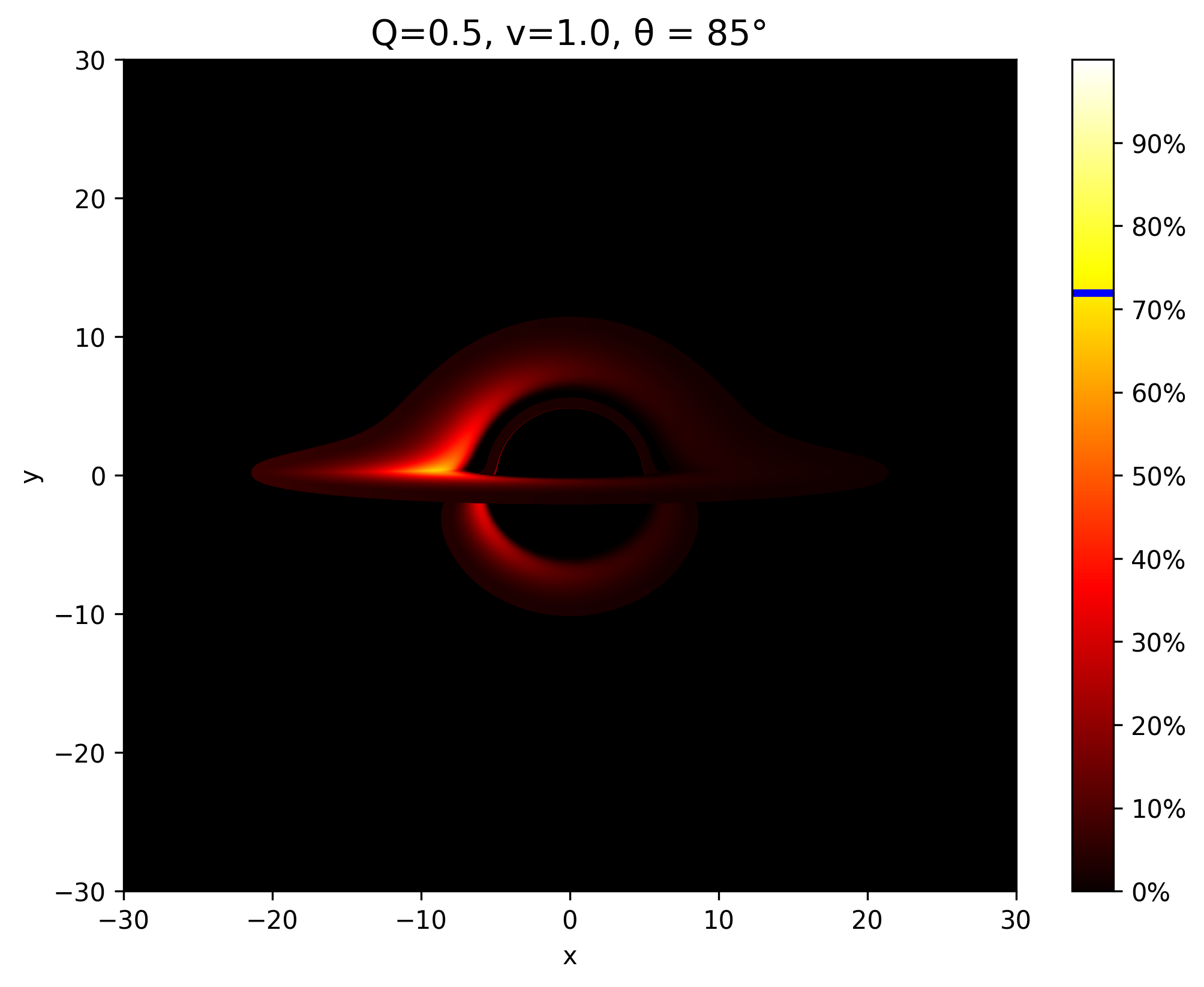}\hspace{-0.2cm}
  \includegraphics[scale=0.35]{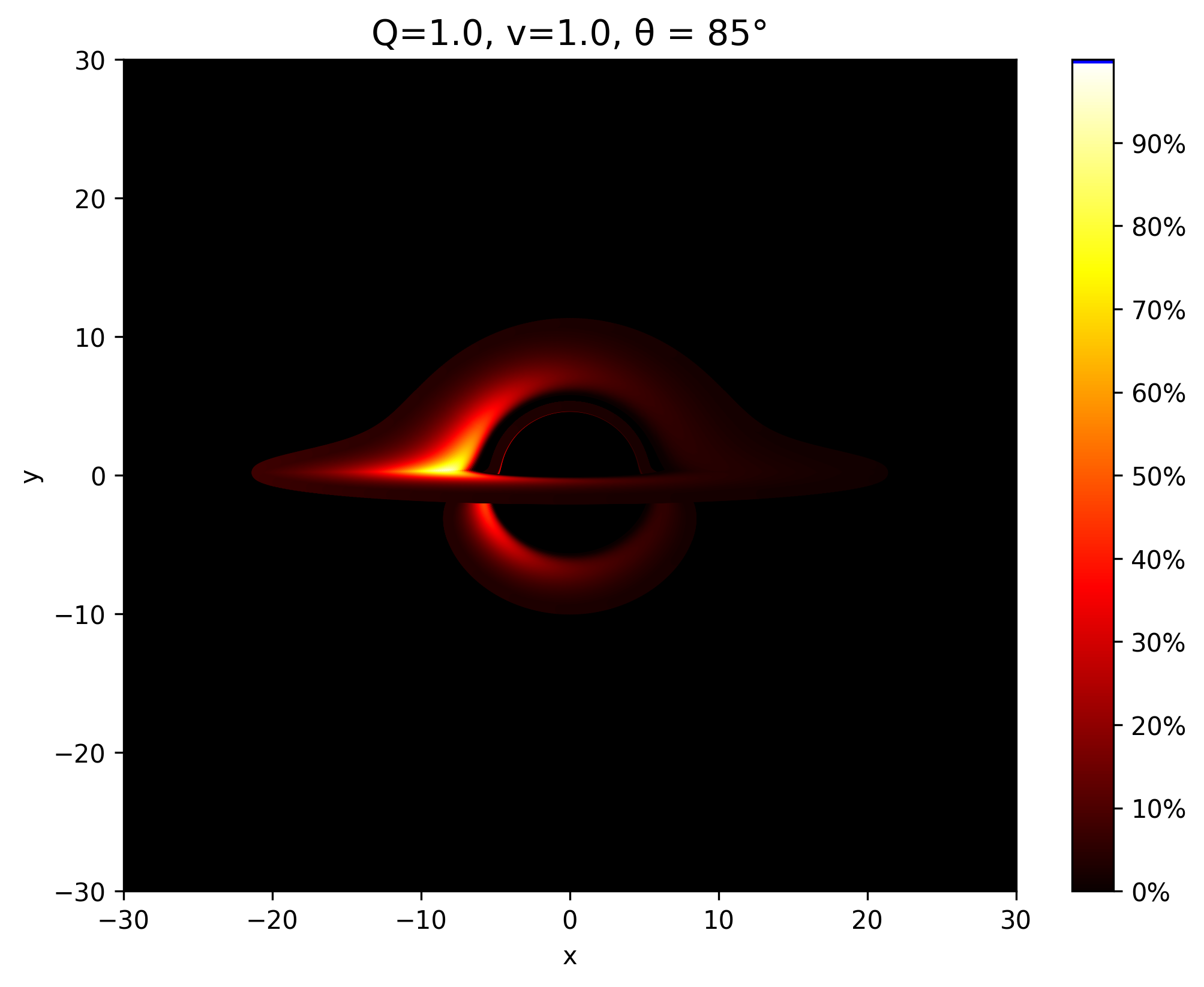}
  \end{tabular}
	\caption{\label{fig:traj2D_Q} 
    Observed flux distributions $F_{\rm obs}$ in the direct and secondary
    images of a thin accretion disk around a ModMax BH for
    $\theta=17^\circ$, $\theta = 53^\circ$, and $\theta = 85^\circ$. The total charge is varied
    as $Q=0$, $Q=0.5$, and $Q=1$, while the nonlinearity parameter is fixed at $v=1$. 
}
\end{figure*}

 \begin{figure*}
\begin{tabular}{ccc}
  \includegraphics[scale=0.35]{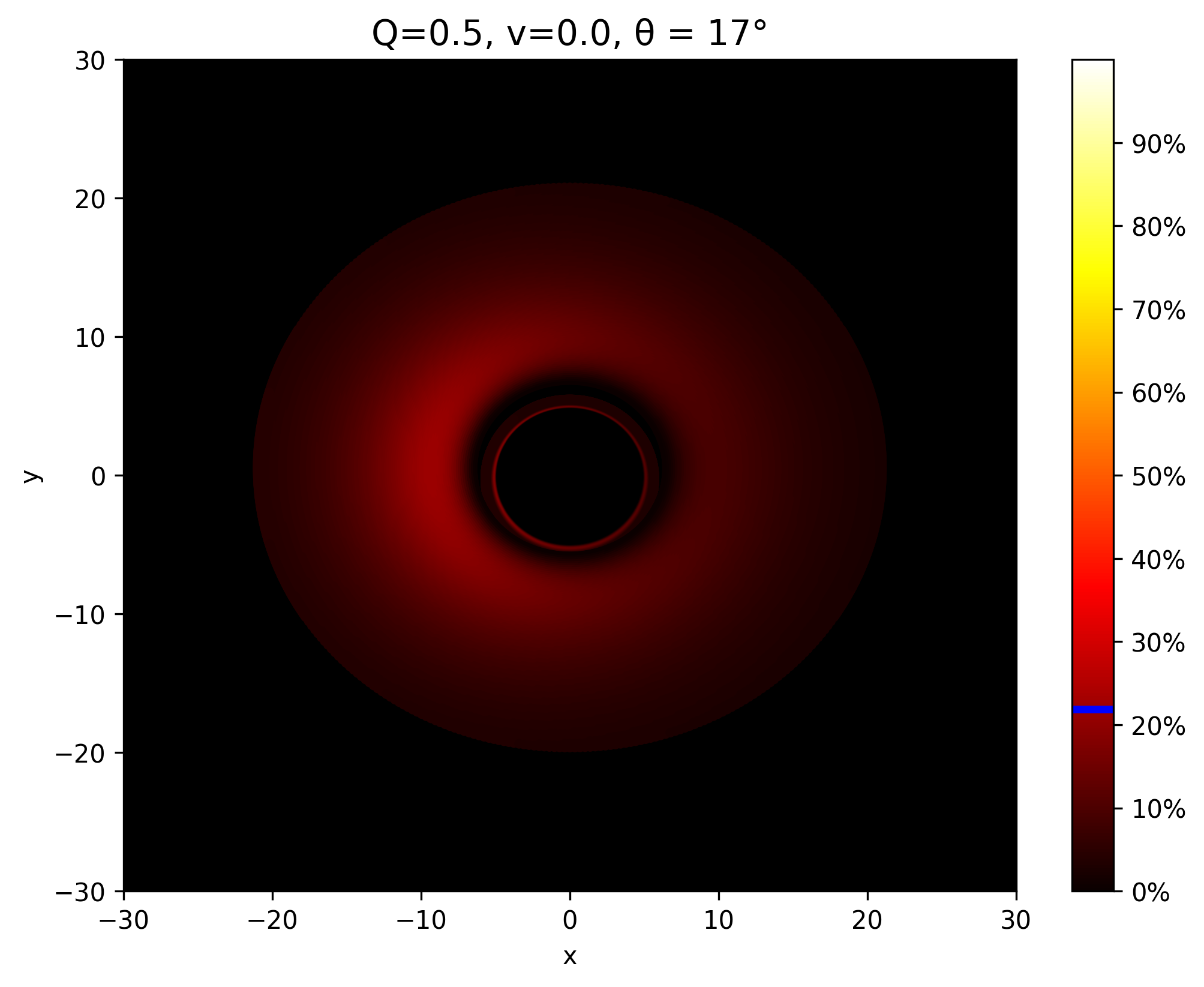}\hspace{-0.2cm}
  \includegraphics[scale=0.35]{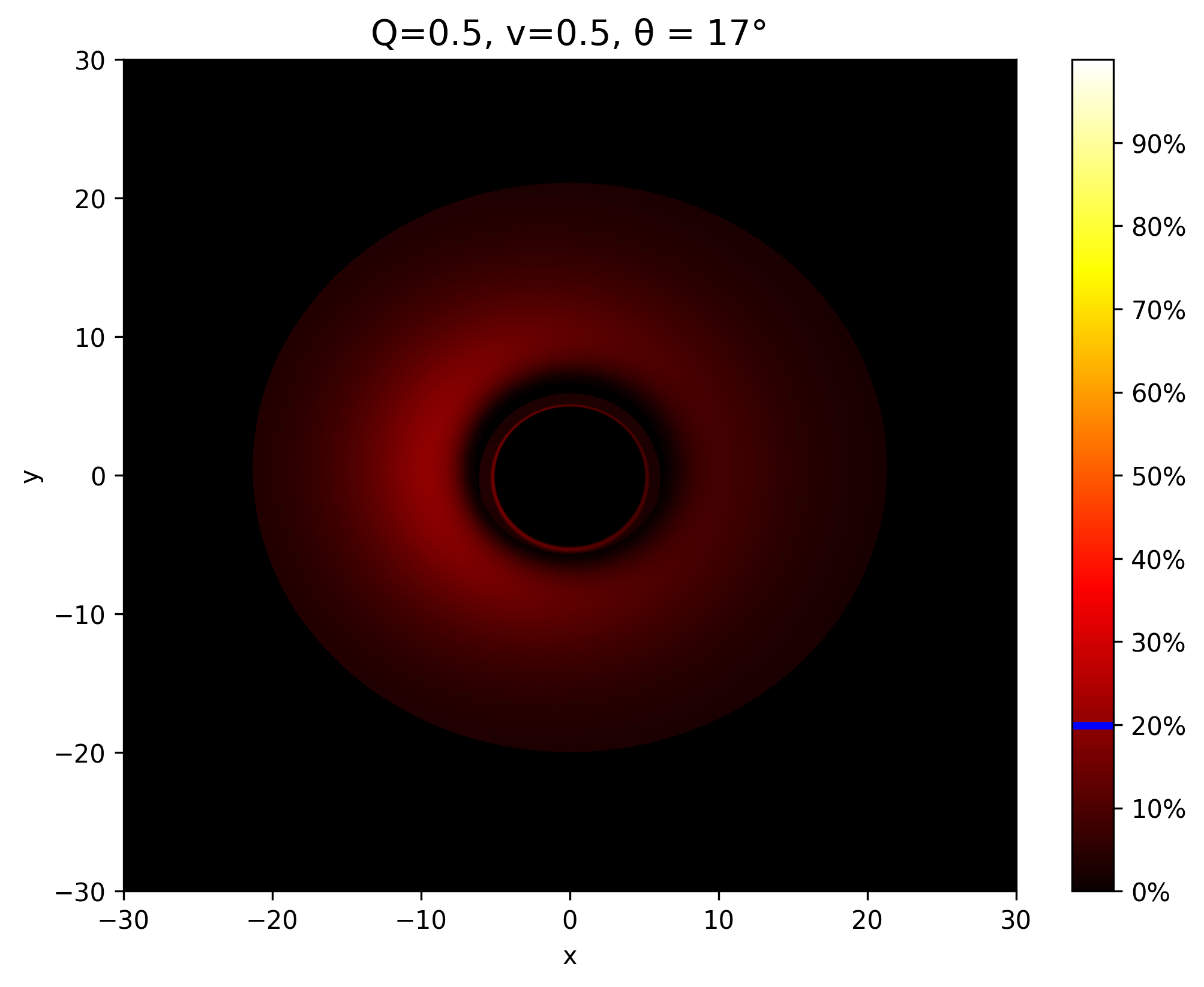}\hspace{-0.2cm}
  \includegraphics[scale=0.35]{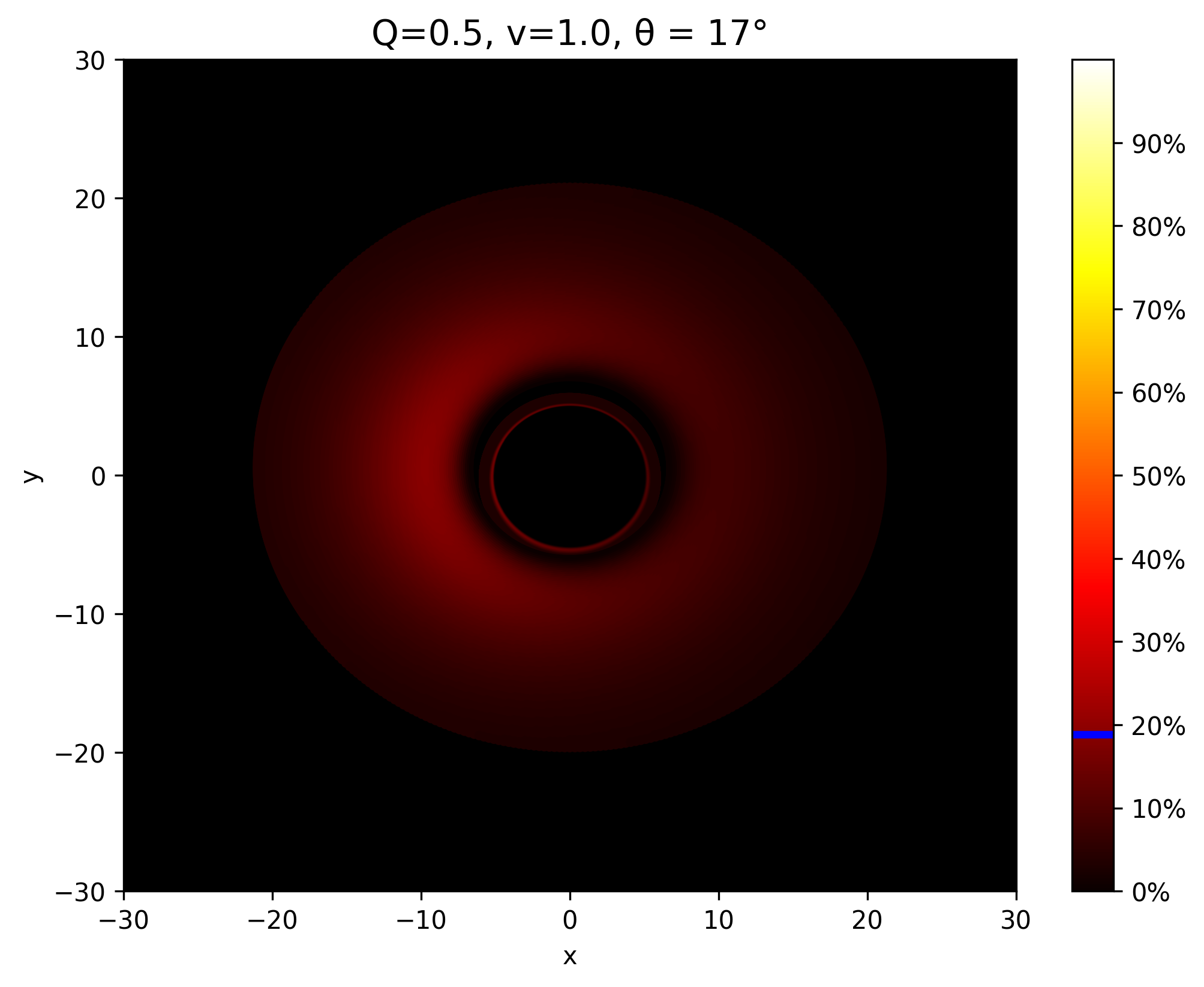}\\
  \includegraphics[scale=0.35]{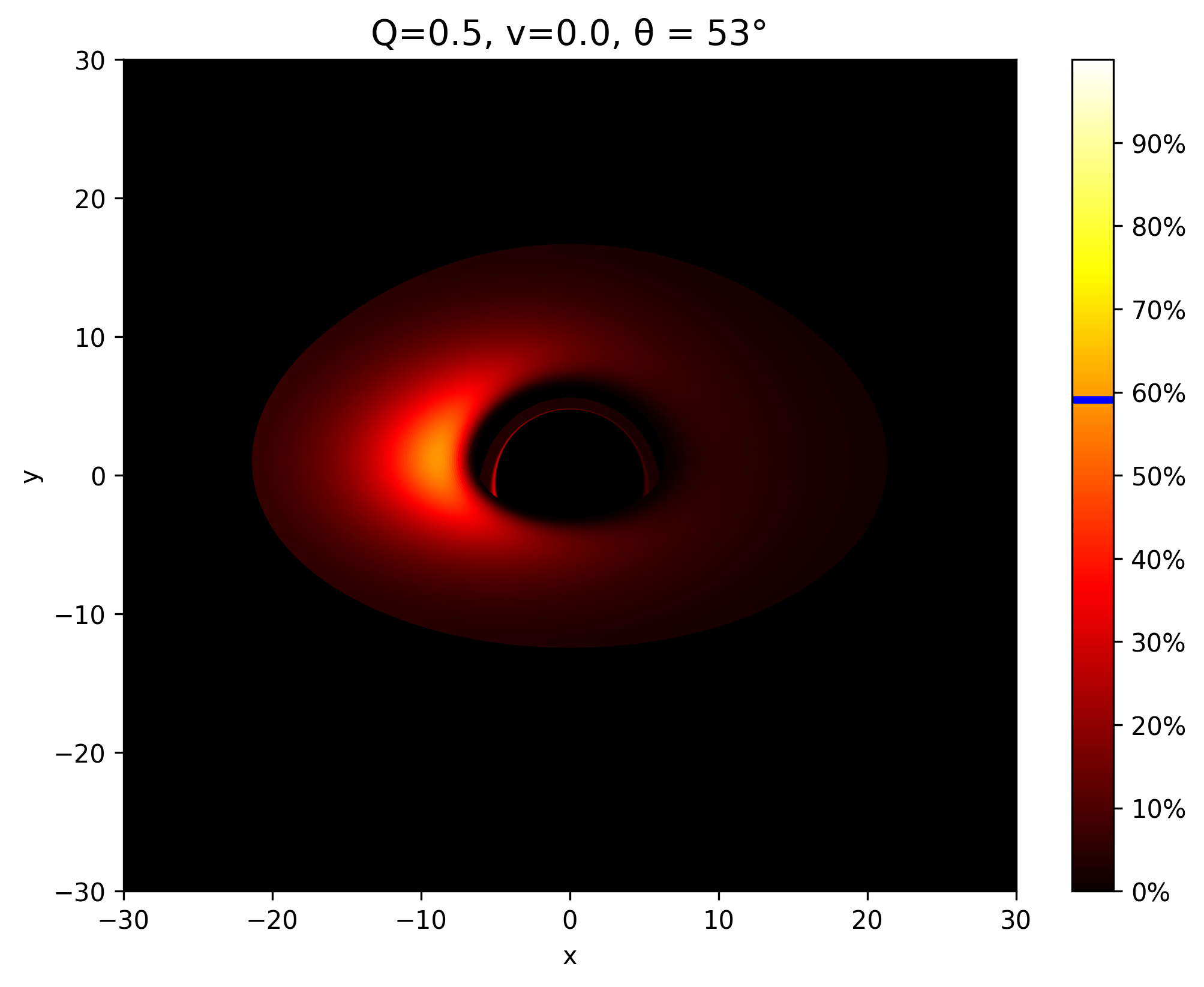}\hspace{-0.2cm}
  \includegraphics[scale=0.35]{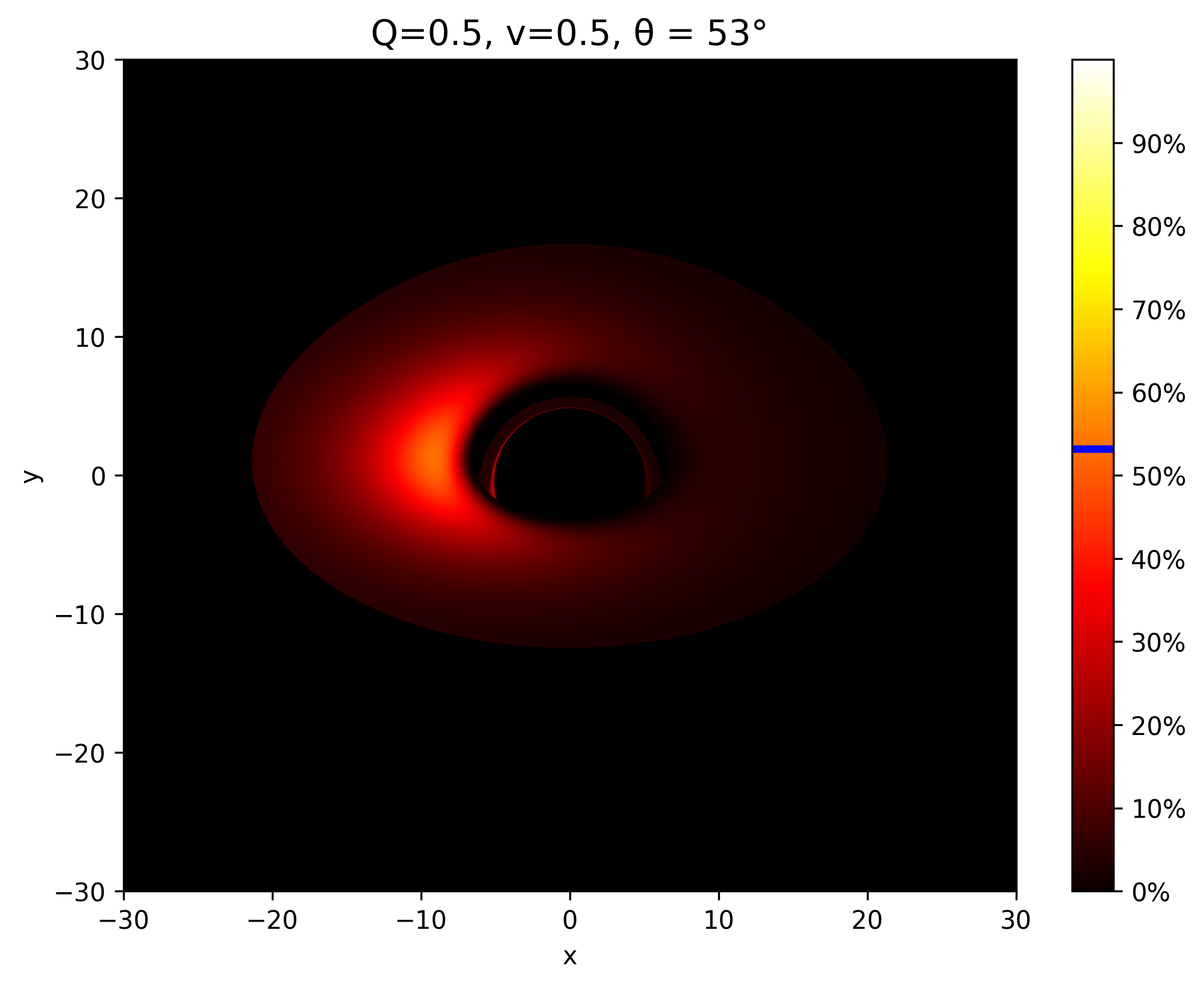}\hspace{-0.2cm}
  \includegraphics[scale=0.35]{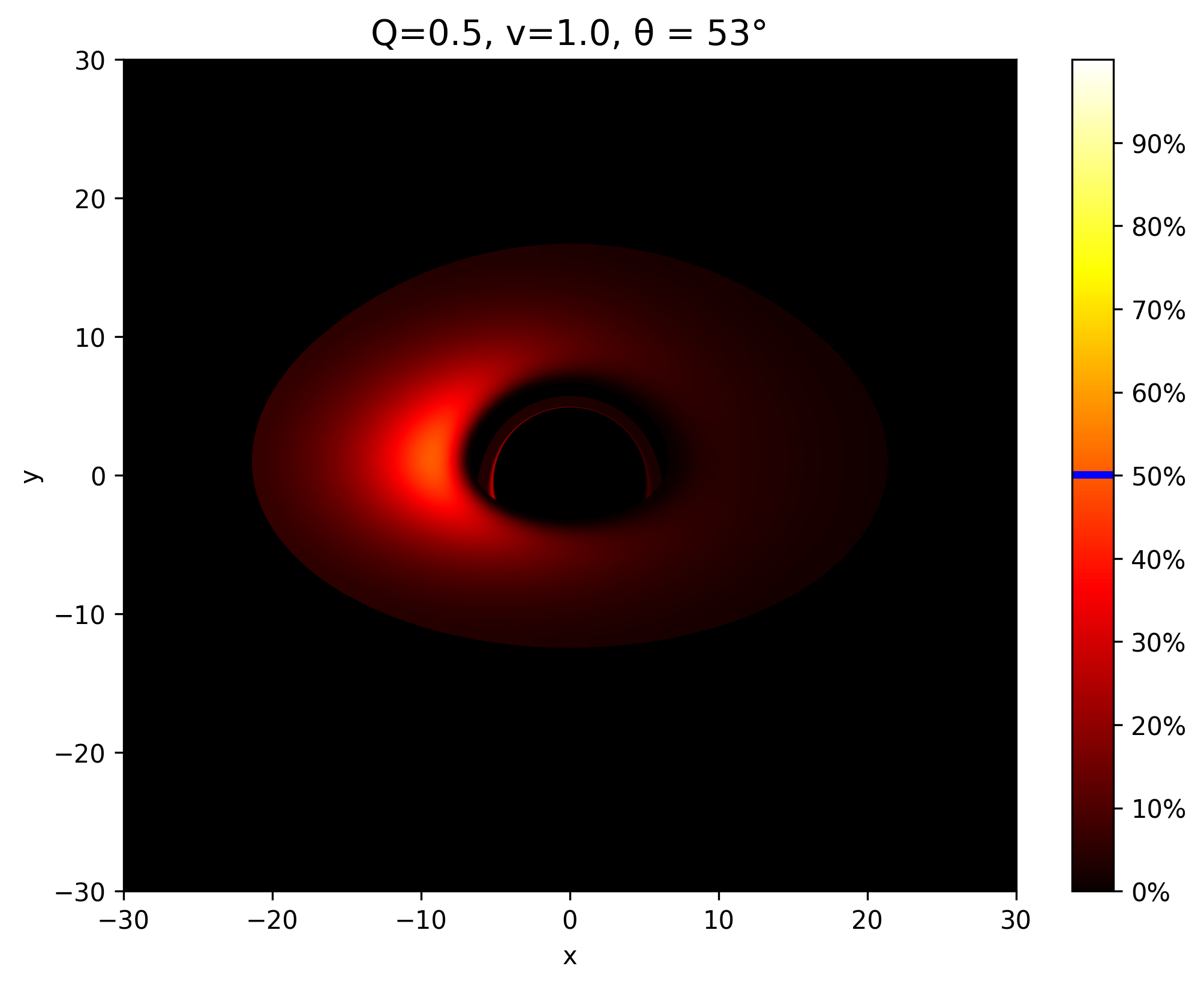}\\
  \includegraphics[scale=0.35]{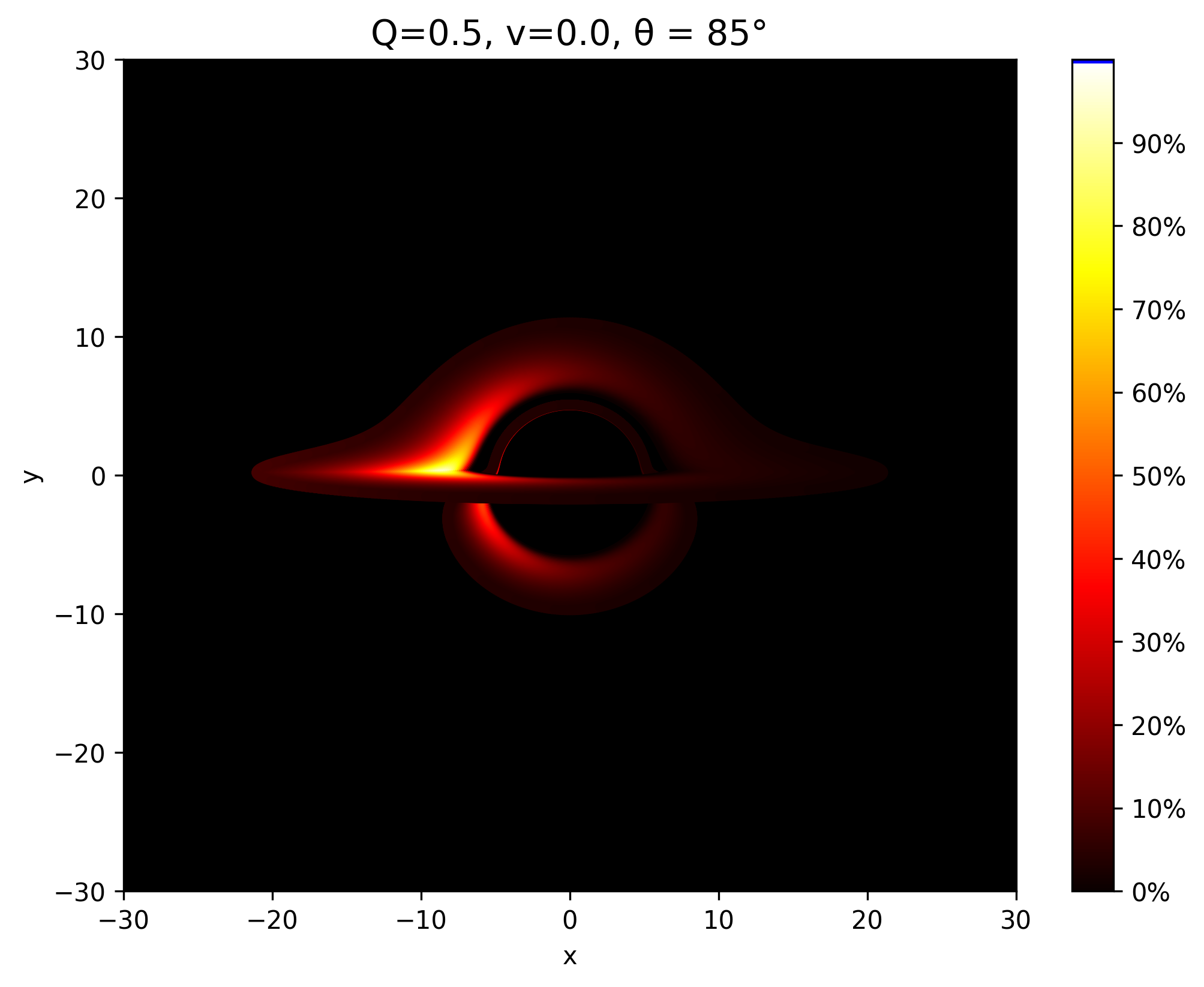}\hspace{-0.2cm}
  \includegraphics[scale=0.35]{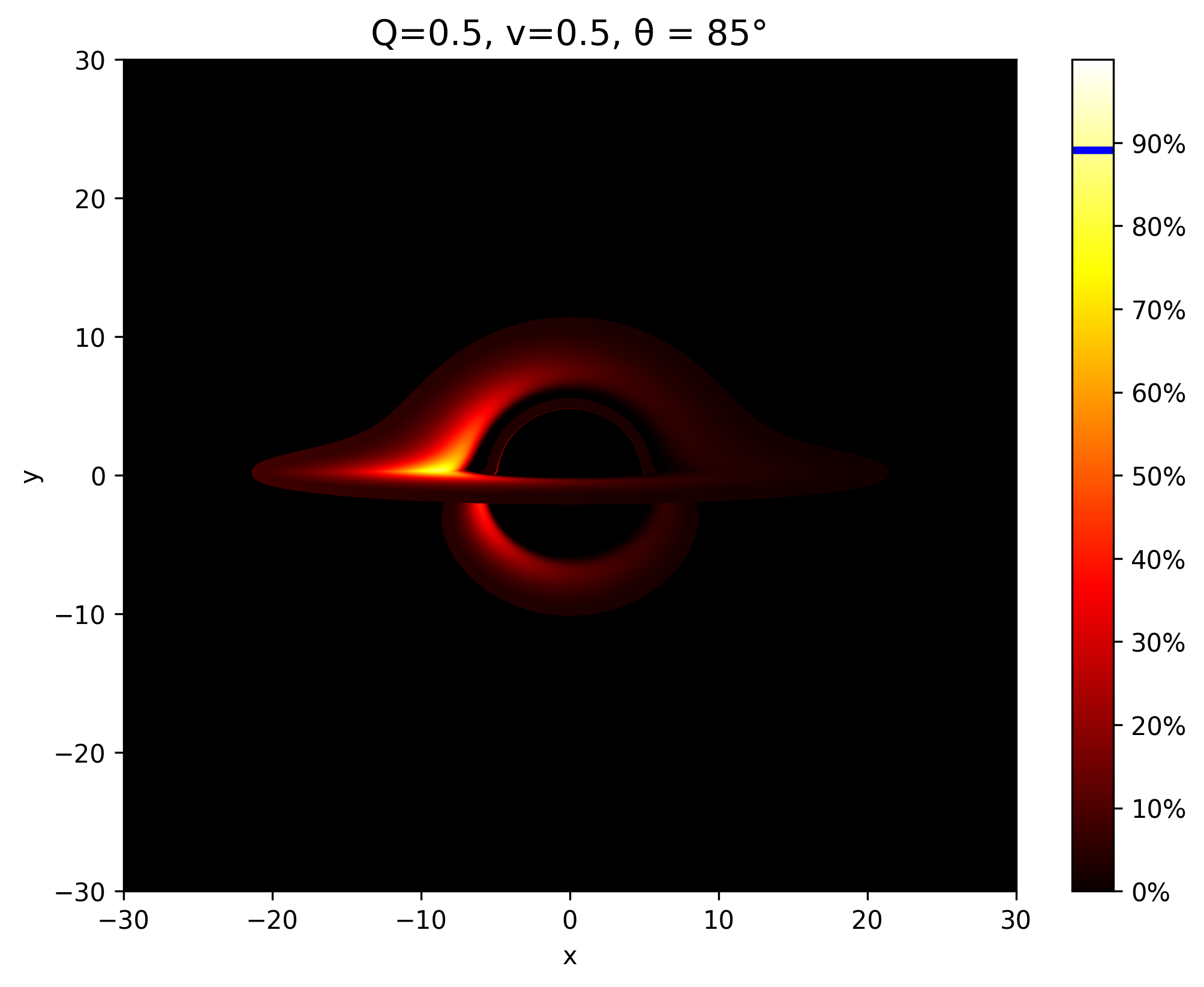}\hspace{-0.2cm}
  \includegraphics[scale=0.35]{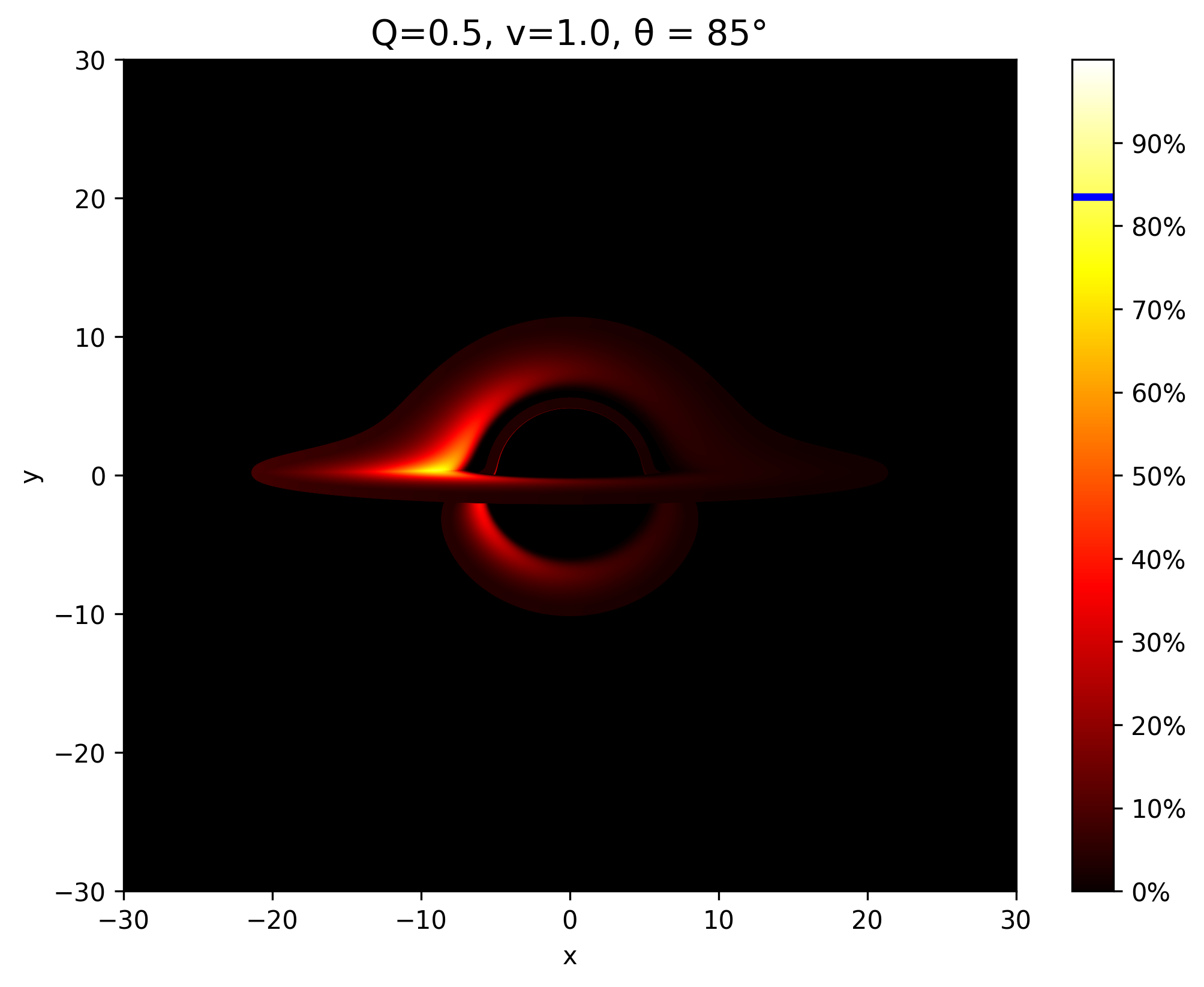}
  \end{tabular}
	\caption{\label{fig:traj2D_theta} 
    Observed flux distributions $F_{\rm obs}$ in the direct and secondary
    images of a thin accretion disk around a ModMax BH for
    $\theta=17^\circ$, $\theta = 53^\circ$, and $\theta = 85^\circ$. The nonlinearity parameter
    is varied as $v=0$, $v = 0.5$, and $v = 1$, while the total charge is fixed at
    $Q=0.5$.
}
\end{figure*}

To explore the observational signatures of the ModMax BH, we analyze
the radiation properties of its accretion disk. Following the Novikov--Thorne thin-disk formalism~\cite{Novikov:1973kta,Shakura:1972te,Thorne:1974ve}, the energy flux emitted from the disk surface takes the form
\begin{equation}
F(r) = -\frac{\dot{M}_0 \Omega_{,r}}{4\pi \sqrt{-g}(E - \Omega L)^2}
\int_{r_{\mathrm{ ISCO}}}^{r} (E - \Omega L)L_{,r}dr\, , 
\label{eq:flux}
\end{equation}
where $\dot{M}_0$ is the mass accretion rate, $g$ denotes the determinant of
the metric, while $E$, $L$, and $\Omega$ represent the specific energy,
specific angular momentum, and angular velocity of particles in circular
orbits, respectively. For circular orbits around a static and spherically symmetric BH, the specific energy, specific angular momentum, and angular velocity
are written as~\cite{Shapiro83,Shaymatov22a}:
\begin{equation}
E = -\frac{g_{tt}}{\sqrt{-g_{tt} - g_{\phi\phi}\Omega^2}}, 
\end{equation}
\begin{equation}
L = \frac{g_{\phi\phi}\Omega}{\sqrt{-g_{tt} - g_{\phi\phi}\Omega^2}}, 
\end{equation}
\begin{equation}
\Omega = \frac{d\phi}{dt} = \sqrt{-\frac{g_{tt,r}}{g_{\phi\phi,r}}}.
\end{equation}

The radiation temperature of the accretion disk is determined from its energy
flux $F(r)$ using the Stefan--Boltzmann law~\cite{Heydari-Fard:2022xhr},
which gives
\begin{equation}
    F(r)=\sigma_{\rm SB}T^4(r)\ .
    \label{eq:T}
\end{equation}

The left panel of Fig.~\ref{fig:fux&temp} shows the radial distribution of
the energy flux $F(r)$ emitted by a thin accretion disk around a ModMax black
hole. Each profile begins at its corresponding innermost stable circular
orbit, $r=r_{\rm ISCO}$. For a fixed nonlinearity parameter $v=1$, increasing
the total charge $Q$ increases the peak flux and shifts its position
slightly toward smaller radii. For a fixed charge $Q=0.75$, increasing $v$
decreases the peak flux and shifts it slightly toward larger radii.

The right panel shows the corresponding temperature distribution $T(r)$.
The temperature distribution shows behavior similar to that of the energy flux: its
maximum increases and moves slightly inward as $Q$ increases, whereas it
decreases and shifts slightly outward as $v$ increases. At large radii, the
differences between the ModMax and Schwarzschild ($Q = 0$) profiles become small.

The flux emitted from the accretion disk is modified by the Doppler effect
and gravitational redshift. For a distant observer, this modification can be
written as
\begin{equation}
F_{\rm obs}=\frac{F(r)}{(1+z)^4}.
\end{equation}
where the redshift factor $z$ is defined as~\cite{Luminet1979}
\begin{equation}
1+z=
\frac{1+\Omega b\sin\theta\cos\alpha}
{\sqrt{-g_{tt}-g_{\phi\phi}\Omega^2}} .
\end{equation}

To investigate the influence of the total charge $Q$ and the nonlinearity parameter $v$ on the distribution of the flux and redshift across the accretion disk around a ModMax BH, we generate simulated disk images for different parameter values and
several observer inclination angles. The images are constructed using the backward ray-tracing method, following the procedure described in Ref.~\cite{Sharipov_2026ChPhC}. For each pixel on the observer's image plane, a photon trajectory is traced backward along null geodesics in the ModMax BH
spacetime. If the ray intersects the accretion disk, 
the local flux $F(r)$ and the observed flux $F_{\rm obs}$ are calculated. The
corresponding flux 
value is then assigned to the pixel on the
observer's image plane. The resulting images are shown in Figs.~\ref{fig:traj2D_Q}, \ref{fig:traj2D_theta}
.
Figs.~\ref{fig:traj2D_Q} and~\ref{fig:traj2D_theta} show the
normalized observed flux distributions $F_{\rm obs}$ across the accretion disk, which extends from $r_{\rm ISCO}$ to $r=25$. The flux maps are normalized with respect to the reference cases $Q=1$, $v=1$ in Fig.~\ref{fig:traj2D_Q} and $Q=0.5$, $v=0$ in Fig.~\ref{fig:traj2D_theta}, respectively. In Fig.~\ref{fig:traj2D_Q}, the nonlinearity parameter is fixed at $v=1$, while the total charge is varied over $Q=0$, $0.5$, and $1$. The observed flux increases with $Q$. For $\theta=85^\circ$, the maximum normalized flux in the Schwarzschild limit ($Q=0$) reaches about $65\%$ of the reference case ($Q=1$, $v=1$, $\theta=85^\circ$). This indicates that the total charge can noticeably affect the disk brightness.
In Fig.~\ref{fig:traj2D_theta}, the total charge is fixed at $Q=0.5$, whereas the nonlinearity parameter is varied over $v=0$, $0.5$, and $1$. The flux maps are normalized with respect to the reference case ($Q=0.5$, $v=0$, $\theta=85^\circ$). In contrast to the effect of $Q$,
increasing $v$ slightly reduces the observed flux. At $\theta=85^\circ$, the maximum normalized flux for $v=1$ is about $85\%$ of that in the reference case, suggesting that the nonlinearity parameter has a weaker effect on the
disk brightness.

\section{Conclusions}
\label{Sec:conclusion}

In this work, we investigated photon motion and the radiative properties of a thin
accretion disk around a ModMax BH and constrained the model parameters
using observational data for M87$^\star$ and Sgr~A$^\star$. We first derived the equations governing null geodesics and determined the
photon-sphere radius and the apparent radius of the BH shadow. It was found that the characteristic radii of the ModMax BH depend on both the total charge $Q$ and the nonlinearity parameter $v$. For fixed $v$, increasing $Q$ reduces the event-horizon, photon-sphere, and shadow radii, whereas for fixed $Q$, increasing $v$ enlarges these radii and shifts them toward their Schwarzschild values.

Using the shadow measurements of M87$^\star$ and Sgr~A$^\star$, we constrained the model parameters $M$, $D$, $Q$, and $v$ through an MCMC analysis. The analysis showed that the obtained values of the BH mass $M$ and distance $D$ were consistent with the observational data. It was found that the best-fit values of $Q$ and $v$ were close to zero, while the observations mainly provided upper limits on these parameters. The strongest constraint on the total charge was obtained from the M87$^\star$ EHT data, $Q<0.391$, whereas the tightest bound on the nonlinearity parameter was obtained from the Sgr~A$^\star$ GRAVITY data, $v<4.153$, both at the 95\% confidence level.

We further studied the radiation properties of a Novikov--Thorne thin disk and generated the corresponding simulated images using the backward ray-tracing method. It was found that the observed flux increases with $Q$, whereas increasing $v$ decreases the disk brightness. For $\theta=85^\circ$, the maximum normalized flux in the Schwarzschild limit was about $65\%$ of that obtained for the ModMax BH with $Q=1$ and $v=1$. For fixed $Q=0.5$, the maximum normalized flux at $v=1$ was found to be about $85\%$ of the reference value at $v=0$.


\acknowledgments

PS would like to acknowledge financial support from the Anusandhan National Research Foundation (ANRF), New Delhi, under grant number CRG/2023/008980.




%
\bibliographystyle{apsrev4-1}  
\bibliography{Ref1,Ref2}

\end{document}